\documentclass[aps,prd,preprint,tightenlines,floatfix]{revtex4}
\usepackage{graphicx}

\newcommand{\nc}{\newcommand}

\nc{\beq}{\begin{equation}}
\nc{\eeq}{\end{equation}}
\nc{\bea}{\begin{eqnarray}}
\nc{\eea}{\end{eqnarray}}
\nc{\n}{\nonumber \\}
\nc{\K}{27 \, \textrm{mK} \, \sqrt{ \frac{1+z}{10} }  \; }
\nc{ \T} {\left( 1 - \frac{T_\gamma}{T_{\rm 0}} \right )}

\begin{document}

\preprint{BI-TP 2009/08 }
\date{September 14, 2009}
\title{Dark matter annihilation and its effect on \\ 
CMB and Hydrogen 21 cm observations}

\author{Aravind Natarajan}
\email{anatarajan@physik.uni-bielefeld.de}
\affiliation{Fakult\"{a}t f\"{u}r Physik, Universit\"{a}t Bielefeld, Universit\"{a}tsstra\ss e 25, Bielefeld 33615, Germany}

\author{Dominik J. Schwarz}
\email{dschwarz@physik.uni-bielefeld.de}
\affiliation{Fakult\"{a}t f\"{u}r Physik, Universit\"{a}t Bielefeld, Universit\"{a}tsstra\ss e 25, Bielefeld 33615, Germany}

\begin{abstract} 
If dark matter is made up of Weakly Interacting Massive Particles, the annihilation of these particles in halos results in energy being released, some of which is absorbed by gas, causing partial ionization and heating. Dark matter annihilation may result in partial ionization and gas heating at high redshifts, even before the formation of the first stars. It is shown that early ionization results in a transfer of power to higher multipoles in the large angle CMB polarization power spectra. Future CMB experiments may be able to place constraints on certain light dark matter models. We also investigate the effect of gas heating on the expected H21 cm power spectrum. Heating by particle annihilation results in a decrease in the amplitude of the H21 cm power spectrum as the gas temperature $T$ becomes comparable to the CMB temperature $T_\gamma$, and then an increase as $T > T_\gamma$. The result is a minimum in the power spectrum at the redshift for which $T \approx T_\gamma$. Only certain models (low particle masses $\sim$ 10 GeV, or favorable halo parameters) show this effect. Within these models, observations of the H21 cm power spectrum at multiple redshifts can help us obtain constraints on dark matter particle and halo properties.
\end{abstract}

\maketitle

\section{Introduction.}

It is widely accepted on the basis of observations of the cosmic microwave background, gravitational lensing, galaxy clusters, rotation curves, etc, that $\sim$ 80\% of the matter in the Universe is in the form of a collisionless, non-baryonic component, commonly referred to as dark matter. Axions and Weakly Interacting Massive Particles (WIMPs) are two of the leading candidates for dark matter. Axion detection experiments include ADMX \cite{admx}, CAST \cite{cast}, Germanium detectors \cite{Ge}, etc. WIMPs may be detected in a variety of ways, both directly and indirectly. Direct detection experiments include DAMA \cite{dama}, CDMS \cite{cdms}, Zeplin \cite{zeplin}, Edelweiss \cite{edelweiss}, etc. Indirect detection experiments include various accelerator and gamma ray search experiments. WIMPs are also relevant to various astrophysical phenomena such as the formation of early stars \cite{stars}, stellar evolution \cite{evol}, etc.

Supersymmetry provides a stable, neutral dark matter candidate called the neutralino. Neutralinos do not decay into standard model particles, but they annihilate in pairs, releasing photons, neutrinos, and charged particles. The energy released by particle annihilation is proportional to the square of the dark matter density and hence, most of the energy is produced by annihilation in dark matter halos. The earliest WIMPy halos form at a redshift $z \sim 60$, with masses $M \sim 10^{-6} M_\odot$ \cite{ghs}, and are thought to be compact and abundant.  Accretion and mergers result in more massive halos, in a bottom-up approach. We restrict ourselves to the case of thermally produced WIMPs with an annihilation cross section determined by the relic density today. 

Observations of the CMB polarization complement information obtained from the temperature anisotropy spectrum. The absence of strong Ly-$\alpha$ absorption lines (the Gunn-Peterson test \cite{gunn}) in the spectra of quasars for $z<6$ implies that the Universe is nearly fully ionized today. The WMAP measured optical depth may be used to place constraints on reionization models. Sudden reionization models with full ionization at redshifts below $z_\ast \sim 11$ result in a peak in the CMB polarization power spectrum at low multipoles (large angles). The polarization power spectrum at large angles is sensitive to the assumed reionization model. Future experiments such as Planck \cite{planck} and CMBPol \cite{pol} may be used to distinguish between different reionization models \cite{haiman}. In this article, we compare the dark matter reionization scenario with the simple case of the sudden reionization model.

 Another powerful probe of the epoch of reionization, and the Universe prior to reionization, is the Hydrogen 21 cm line. Observations of the fluctuations in the H21 cm differential brightness temperature serve as a probe of the Universe at redshifts $z < 200$ \cite{Hydrogen}. Before the formation of the first luminous sources, the H21 cm signal is expected to be negligible at redshifts $z\lesssim30$. This scenario is altered if dark matter annihilation provides a source of heating. Future experiments such as LOFAR \cite{lofar} and the Square Kilometer Array \cite{ska} aim to study the fluctuations in the H21 cm temperature, and provide important information regarding reionization and the high redshift Universe. Several authors have studied the impact of particle annihilation on reionization \cite{ion,ripa,chuzhoy,previous_work,dom,hooper} and the Hydrogen 21 cm signal \cite{chuzhoy,furl1,furl_review,H21}. In this article, we investigate possible signatures of dark matter annihilation in the CMB and Hydrogen 21 cm power spectra at redshifts before the formation of the first luminous objects, i.e. between $z \sim 60$ and $z \sim 10$.

In Section II, we consider dark matter halos fitted with an NFW profile, and calculate the luminosity of the halos. We then compute the energy absorbed by gas at any redshift $z$ due to particle annihilation in dark matter halos at redshifts $z' > z$.  Some of the absorbed energy results in ionization and heating. We include the effect of gas heating in computing the evolution of the ionized fraction, thus improving upon the calculation in \cite{previous_work}.

In Section III, we calculate the optical depth due to scattering of free electrons with CMB photons. Early ionization results in more scattering of free electrons with CMB photons, and hence a larger polarization signal. However, the partial ionization at high redshifts requires a smaller ionization at low redshifts in order to keep the optical depth constant. Thus the large angle polarization power spectrum is modified. The effect is small, but detectable for small particle masses ($m_\chi \sim 2$ GeV), and favorable halo parameters. 

In Section IV, we study the Hydrogen 21 cm differential brightness temperature. We show that the increase in gas temperature due to particle annihilation at high redshifts can have an observable effect on the power spectrum of fluctuations, for certain light dark matter models ($m_\chi \sim 10$ GeV). As the gas temperature $T$ increases and becomes comparable to the CMB temperature $T_\gamma$, the amplitude of fluctuations in the  H21 cm differential brightness temperature decreases and reaches a minimum. The amplitude of fluctuations then increases as $T > T_\gamma$. We compare the power spectrum to the standard scenario that includes no dark matter heating. Finally, we present our conclusions.

\section{Particle Annihilation in dark matter halos.}

Consider dark matter halos fitted with a Navarro-Frenk-White(NFW) density profile \cite{nfw}:
\beq
\rho(r) = \frac{\rho_{\rm s}}   { (r/r_{\rm s}) \left[1 + r/r_{\rm s}\right ]^2}.
\label{nfw}
\eeq
$\rho_{\rm s}$ and $r_{\rm s}$ are constants. We define the concentration parameter $c_{\rm 200} = r_{\rm 200} / r_{\rm s}$. $r_{\rm 200}$ is the radius at which the enclosed mean density equals 200 times the cosmological average measured at the redshift $z_{\rm F}(M)$ at which the halo formed:
\beq
\frac{3M}{4 \pi r^3_{\rm 200}} = \bar\rho(M) = 200 \rho_{\rm c} \Omega_{\rm m} \left[ 1 + z_{\rm F}(M) \right ]^3.
\eeq
$M = M(r_{\rm 200})$ is the mass of the halo. We set the mass in dark matter to be $M_{\rm dm} = (\Omega_{\rm dm} / \Omega_{\rm m})  \; M$. $\rho_{\rm c}$ is the critical density, and $\Omega_{\rm m}$ is the matter fraction. $c_{\rm 200}$ is thus measured at the time of halo formation. In this article, we treat $c_{\rm 200}$ as a free parameter,  independent of the halo mass. Particle annihilation in dark matter halos results in a luminosity (energy per unit time) per photon energy = $dL/dE_\gamma$ given by \cite{dark_matter}:
\bea
\frac{dL}{dE_\gamma} &=& \frac{dN_\gamma}{dE_\gamma} \, E_\gamma \times \frac{\langle \sigma_{\rm a} v \rangle}{2\,m^2_\chi}  \, \int dr \, 4 \pi r^2 \, \rho^2(r) \n
&=& \frac{dN_\gamma}{dE_\gamma} \, E_\gamma \; \frac{\langle \sigma_{\rm a} v \rangle}{2\,m^2_\chi}   \; \frac{M \, \bar\rho}{3} \left( \frac{\Omega_{\rm dm}}{\Omega_{\rm m}} \right )^2 \, f(c_{\rm 200}).
\label{lum_equation}
\eea
$m_\chi$ is the particle mass. $dN_\gamma/dE_\gamma = m^{-1}_\chi \, dN_\gamma/dx$ is the number of photons released per annihilation per photon energy. Following \cite{pheno} for mixed gaugino-higgsino models , we choose $dN_{\rm \gamma}/dx$ to have the form  $a\,e^{-bx} / x^{1.5}$, where $a$ and $b$ are constants for a particular annihilation channel. Averaging over the channels considered in \cite{pheno}, we find $a = 0.9, b = 9.56$.  $\langle \sigma_{\rm a} v \rangle$ is the annihilation cross section of the WIMPs times the relative velocity, averaged over the velocity distribution. We set $\langle \sigma_{\rm a} v \rangle = 3 \times 10^{-26}$ cm$^3$/s, assumed independent of $v$ (i.e. $s$-wave annihilation).  For the NFW density profile Eq. (\ref{nfw}) , we have
\beq
f(c_{\rm 200}) = \frac{ c^3_{\rm 200} \, \left[ 1 - (1+c_{\rm 200})^{-3}  \right ] }{ 3 \left[ \ln (1+c_{\rm 200}) - c_{\rm 200}(1+c_{\rm 200})^{-1} \right ]^2 }.
\eeq
$f(c_{\rm 200})$ takes values 18, 45, 150, and 610, for $c_{\rm 200}$ = 2.5, 5, 10, and 20 respectively. In this paper, we restrict ourselves to the density profile of Eq. [\ref{nfw}]. The luminosity can be much larger for halo profiles that are more cuspy. For density profiles with $\rho$ falling off faster than $r^{-1.5}$, the luminosity is divergent unless a lower cutoff radius is imposed \cite{previous_work}. 

\subsection{Interaction of photons with gas atoms.}
Consider dark matter annihilation at a location $s'$. Some of the photons released at $s'$ travel to a location $s$ where they scatter with gas atoms, resulting in ionization and heating. The probability of scattering between $s$ and $s+\delta s$ is given by $ \delta s \, \sigma(E_\gamma) \, n \,  [1+z(s)]^3$ where
\beq
\delta s = c \, \delta t = -c \, \delta z / \left[ (1+z) H(z) \right ].
\eeq
$n (1+z)^3$ is the physical number density of atoms at $z$. The fraction of photons that reach $s$ having been released at $s'$ is
\bea
\kappa(s,s',E_\gamma) = \exp \left[ - \sigma(E_\gamma) \, n  \,  \int_{s'}^s \, ds'' \left[1 + z(s'') \right]^3 \right ] \n
\kappa(z,z',E_\gamma) = \exp - \frac{2A}{3} f_\sigma(E_\gamma) \left[ (1+z')^{3/2} - (1+z)^{3/2} \right ].
\eea
$A$ is the dimensionless quantity
\beq
A = \frac{c \, \sigma_{\rm T}  n}{H_{\rm 0} \sqrt{\Omega_{\rm m}}} \approx 3.5 \times 10^{-3},
\eeq
and $f_\sigma(E_\gamma)$ is the energy dependent correction to the Thomson cross section $\sigma_{\rm T}$ (see for e.g., \cite{peskin}):
\beq
f_\sigma(y) = \frac{3}{8}  \left[ \frac{2(1+y)}{(1+2y)^2} + \frac{\log(1 + 2y)}{y} - \frac{2}{1+2y} + \frac{2(1+y)^2}{y^2(1+2y)} - \frac{2(1+y) \log(1+2y)}{y^3} + \frac{2}{y^2} \right ],
\eeq
where $y = E_\gamma / m_{\rm e}$ and $m_{\rm e}$ is the electron mass. For energies $E_\gamma \gg 13.6$ eV, we may ignore the bound structure of the atom, and consider scattering of photons with electrons. 

The comoving number density of dark matter halos at a redshift $z$ is given by the Press-Schechter formula \cite{press_schechter}
\beq
\frac{dn_{\rm h}}{dM}(z,M) = \sqrt{\frac{2}{\pi}} \, \frac{\rho_{\rm c} \Omega_{\rm m}}{M} \, \frac{\delta_{\rm c} \, (1+z)}{\sigma^2_{\rm h}} \, \frac{d\sigma_{\rm h}}{dM} \, \exp \left[ -\frac{\delta^2_{\rm c} (1+z)^2}{2 \sigma^2_{\rm h}} \right ].
\label{press_schechter}
\eeq
$\sigma^2_{\rm h}$ is the variance of halo fluctuations (see for e.g., \cite{cooray}):
\bea
\sigma^2_{\rm h}(R) &=& \int \frac{dk}{k} \, \frac{k^3 P(k)}{2 \pi^2} \, \left | W(kR) \right |^2  \nonumber \\
W(kR) &=& \frac{3}{(kR)^3} \, \left[ \sin kR - (kR) \cos kR \right ].
\eea
We use the code of Eisenstein and Hu \cite{eisenstein} to compute the matter power spectrum $P(k)$. $P(k)$ is normalized so that $\sigma_{\rm 8} = \sigma_{\rm h}(R = 8 h^{-1} \textrm{Mpc})= 0.8$.  $\delta_{\rm c}$ is set to 1.28 \cite{barkana}. Assuming density fluctuations grow as $(1+z)^{-1}$, the matter power spectrum at redshift $z$ is $P(z,k) = P(k)/(1+z)^2$. 

\begin{figure}[!ht]
\begin{center}
\scalebox{0.9}{\includegraphics{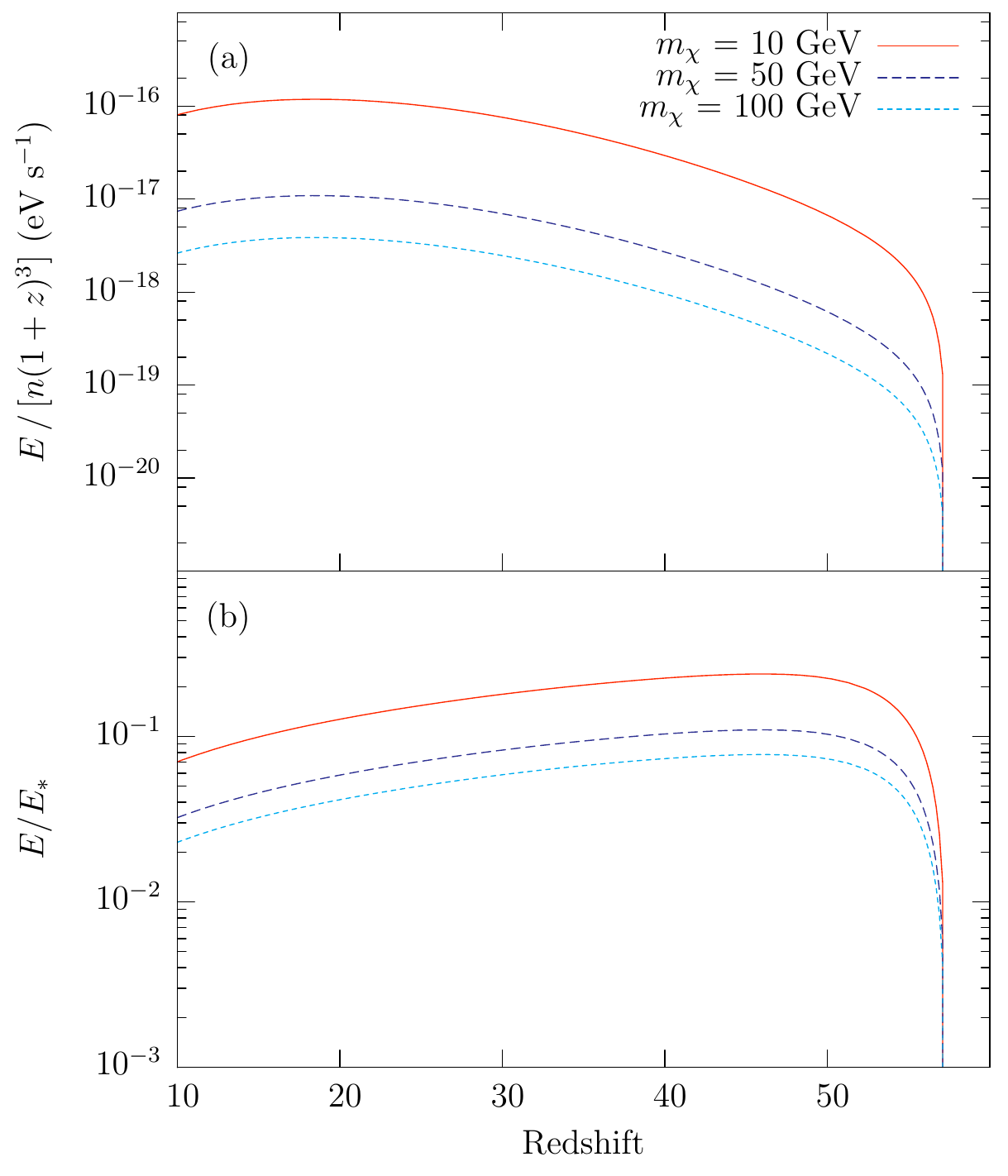}}
\end{center}
\caption{  (a) shows the average absorbed energy per atom per unit time ($E/n(1+z)^3$) plotted as a function of redshift, for particle masses $m_\chi = 10,50$, and 100 GeV, with $c_{\rm 200} = 10$. (b) shows $E/E_\ast$. Peak absorption $\sim 20 \%$ for $m_\chi = 10$ GeV, and $\sim 6 \%$ for $m_\chi = 100$ GeV. \label{fig1} }
\end{figure}

\begin{figure}[!ht]
\begin{center}
\scalebox{0.9}{\includegraphics{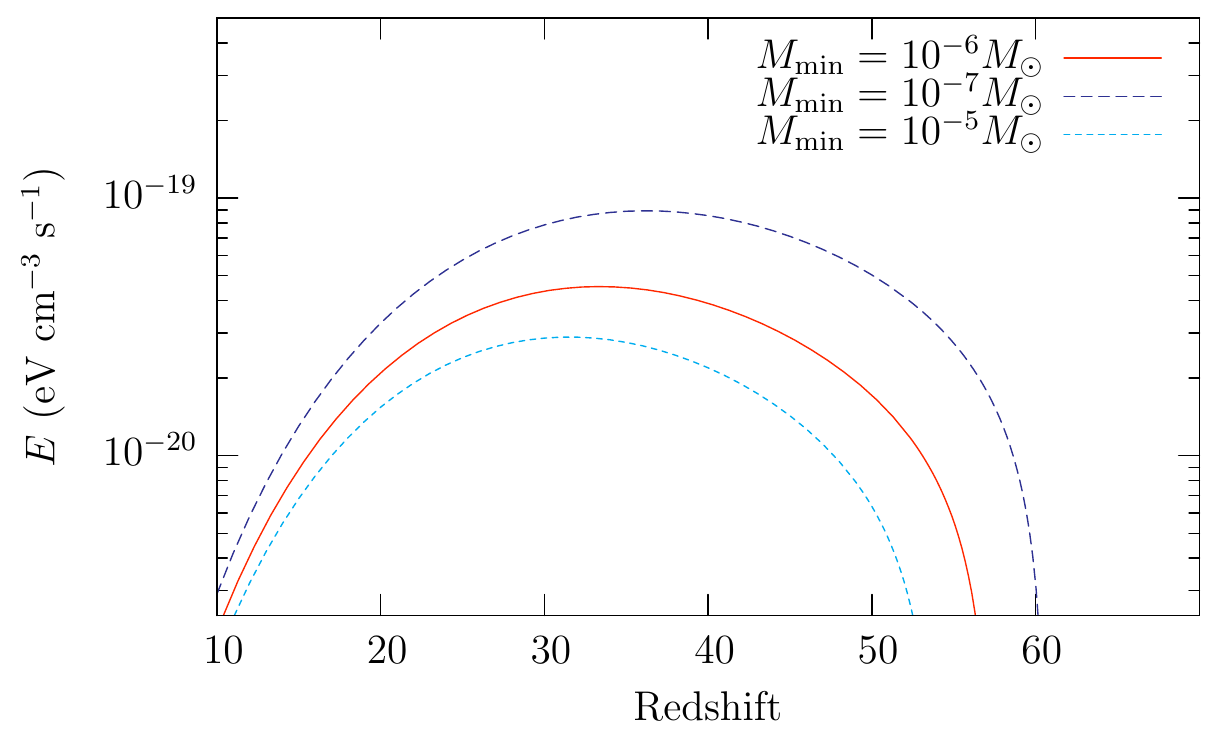}}
\end{center}
\caption{  Different choices of the minimum halo mass $M_{\rm min}$. $E$ is small at late times because of the low halo and gas densities, and at very early times because halos are rare. The formation redshift is larger for smaller $M_{\rm min}$. The particle mass $m_\chi = 50$ GeV.  \label{fig2} }
\end{figure}

The energy absorbed by gas at a redshift $z$ is $E(z,M_{\rm min},m_\chi,c_{\rm 200})$ given by (see \cite{previous_work} for details)
\bea
E = \frac{ 25 A \, \langle \sigma_{\rm a} v \rangle }{m_\chi} \, \frac{ f(c_{\rm 200}) }{ f(c_{\rm 200} = 10) } \left( \frac{\Omega_{\rm dm}}{\Omega_{\rm m}} \right )^2 \,  (1+z)^5  \int_{z_{\rm F}}^z \, -dz' \, (1+z')^{-1/2}  \n
\times  \int_0^1 dx \frac{a e^{-bx}}{\sqrt{x}} f_\sigma(m_\chi x) \, \kappa(z,z',m_\chi x) \; \int_{M_{\rm min}}^\infty \, dM \, \frac{dn_{\rm h}}{dM} (z',M) \,  \bar \rho(M) \, M.
\label{E_equation}
\eea
$z_{\rm F} = z_{\rm F}(M_{\rm min})$ is that redshift at which $\sigma_{\rm h} (M_{\rm min},z_{\rm F}) = 1$ provided $z_{\rm F} > z$ \cite{ghs}. If not, we set $z_{\rm F} = z$. We use the form given by \cite{ghs} to estimate $z_{\rm F}(M)$.  The energy absorbed by the gas contributes to ionization as well as heating. Monte Carlo simulations \cite{monte} suggest that about a third of the absorbed energy goes into ionization, while a third goes into heating. We therefore set the efficiency factors $\eta_{\rm ion} = \eta_{\rm heat} = 0.3$. Fig. \ref{fig1}(a) shows the average energy absorbed per atom per unit time ($E / n (1+z)^3$), for particle masses $m_\chi = 10,50$, and 100 GeV. 

It is instructive to compare $E$ to the simple approximation $E_\ast$:
\beq
E_\ast(z) = \int dM \, (1+z)^3 \frac{dn_{\rm h}}{dM}(z,M) \; \int dE_\gamma \, \frac{dL}{dE_\gamma}(E_\gamma,M),
\label{E_star}
\eeq
which neglects the effect of propagation of photons, and unlike $E$, is a strictly local function of $z$. Fig. \ref{fig1}(b) compares $E$ and $E_\ast$. We see that for light dark matter particles ($m_\chi \approx 10$ GeV), about 20\% of the energy released is absorbed at high redshifts, falling to $\approx 10 \%$ at lower redshifts. This is in agreement with the somewhat larger absorption fraction obtained by \cite{ripa} at $z=50$, for very light dark matter with mass $m_\chi = 10$ MeV. Assuming the efficiency factors $\eta_{\rm ion} = \eta_{\rm heat} = 0.3$, we find that $ \lesssim 0.3\times 0.2 = 6$\% of the generated energy goes into ionization, and an equal amount into heating. The maximum absorbed fraction is lower ($\approx 6 \%$) for 100 GeV dark matter particles, with a useful fraction $\lesssim 2 \%$ each for ionization and heating.  Fig. \ref{fig2} shows the effect of varying the assumed minimum halo mass $M_{\rm min}$. The minimum halo mass is set by the free streaming scale which depends on the particle physics of the model. For WIMP dark matter, we expect $M_{\rm min} \approx 10^{-6} M_\odot$ \cite{ghs}. We set $M_{\rm min}$ to $10^{-6} M_\odot$ in the remainder of this article. We now calculate the evolution of the ionized fraction and the gas temperature.

\subsection{Ionization of gas.}

The number of ionizations per unit volume per unit time, at a redshift $z$ is given by
\beq
I(z) = \mu \, (1-x_{\rm ion}) \, \eta_{\rm ion} E(z).
\eeq
$\mu$ is the inverse ionization potential
\beq
\mu = \left[ \frac{0.76}{0.82} \; \frac{1}{13.6 \,\textrm{eV}} + \frac{0.06}{0.82} \; \frac{1}{24.6 \,\textrm{eV}} \right ]  \approx 0.07 \; \rm{eV}^{-1},
\label{mu}
\eeq
where we assumed $76 \%$ H and $24 \%$ He by mass. Only singly ionized Helium is considered in Eq. (\ref{mu}), and $x_{\rm ion}$ is the ionized fraction. The number of recombinations per unit volume, per unit time at $z$ is given by
\beq
R(z) =  n^2\, x^2_{\rm ion}(z) (1+z)^6 \alpha(T).
\eeq
$T$ is the gas temperature and $\alpha$ is the rate coefficient
\beq
\alpha(T) = \left[ \frac{0.76}{0.82} \, \alpha_{\rm H}(T) + \frac{0.06}{0.82} \, \alpha_{\rm He}(T) \right ],
\eeq
with $\alpha_{\rm H}$ and $\alpha_{\rm He}$ given by \cite{abel}
\bea
\alpha_{\rm H} &\approx& 3.746 \times 10^{-13} (T/\rm{eV})^{-0.724} \;\; \textrm{cm}^3 \, \textrm{s}^{-1} \n
\alpha_{\rm He} &\approx& 3.925 \times 10^{-13} (T/\rm{eV})^{-0.6353} \;\; \textrm{cm}^3 \, \textrm{s}^{-1}.
\eea
$x_{\rm ion}$ is obtained by solving the equation
\beq
I(z) - R(z) = -n H_{\rm 0} \sqrt{\Omega_{\rm m}} (1+z)^{11/2} \frac{dx_{\rm ion}}{dz},
\label{ion}
\eeq
where we used
\beq
dz = - dt \, H_{\rm 0} \sqrt{\Omega_{\rm m}} (1+z)^{5/2} \, \left[ 1 + \frac{\Omega_\Lambda}{\Omega_{\rm m}} (1+z)^{-3} \right ]^{1/2},
\eeq
and neglected the effect of dark energy for $z > 10$. In order to solve Eq. (\ref{ion}), we need to obtain an expression for $T(z)$.

\subsection{Heating.}

In the absence of heating, the gas temperature falls off $\propto (1+z)^2$. Dark matter annihilation results in a heating term $(dT/dz)_{\rm heat}$
\beq
\eta_{\rm heat} \, E(z) = -\frac{3}{2} \,n k_{\rm b} H_{\rm 0} \sqrt{\Omega_{\rm m}} (1+z)^{11/2} \, \left( \frac{dT}{dz} \right )_{\rm heat},
\eeq
where $k_{\rm b}$ is Boltzmann's constant. In the presence of free electrons, energy is transferred between the photons and the gas by Thomson scattering, resulting in a coupling term \cite{coupling_term,furl1}
\bea
-H_{\rm 0} \sqrt{\Omega_{\rm m}} (1+z)^{5/2} \left(\frac{dT}{dz}\right )_{\rm coup} &=& \frac{8 a \sigma_{\rm T}  T^4_\gamma(z)  }{3  m_{\rm e} c} \frac{x_{\rm ion}}{1 + x_{\rm ion} + f_{\rm He} } \left [ T_\gamma(z) - T(z) \right ]\n
&=& \frac{T_\gamma(z) - T(z)}{t_{\rm c}(z)} \, \frac{x_{\rm ion}}{1 + x_{\rm ion} + f_{\rm He} },
\eea
where $a$ is the radiation constant, and $f_{\rm He}$ is the Helium fraction. $t_{\rm c}(z) = 3 m_{\rm e} c / 8 a \sigma_{\rm T} T^4_\gamma(z) \approx$ 1.44 Myr $[30/(1+z)]^4$. This coupling increases the gas temperature when $T$ is smaller than the CMB temperature $T_\gamma$ and decreases it when $T > T_\gamma$. The evolution of the gas temperature thus follows the equation
\beq
\frac{dT}{dz} = \frac{2T}{1+z} - \frac{1}{H_{\rm 0} \sqrt{\Omega_{\rm m}} (1+z)^{5/2}} \left[ \frac{2 \eta_{\rm heat}  E }{3  k_{\rm b}  n (1+z)^3 }  + \, \frac{x_{\rm ion}}{1+x_{\rm ion}+f_{\rm He}} \, \frac{T_\gamma - T}{  \; t_{\rm c}}
\label{T} \right ].
\eeq
Solving Eq. (\ref{T}) together with Eq. (\ref{ion}) gives $T(z)$ and $x_{\rm ion}(z)$.

\begin{figure}[!ht]
\begin{center}
\scalebox{0.8}{\includegraphics{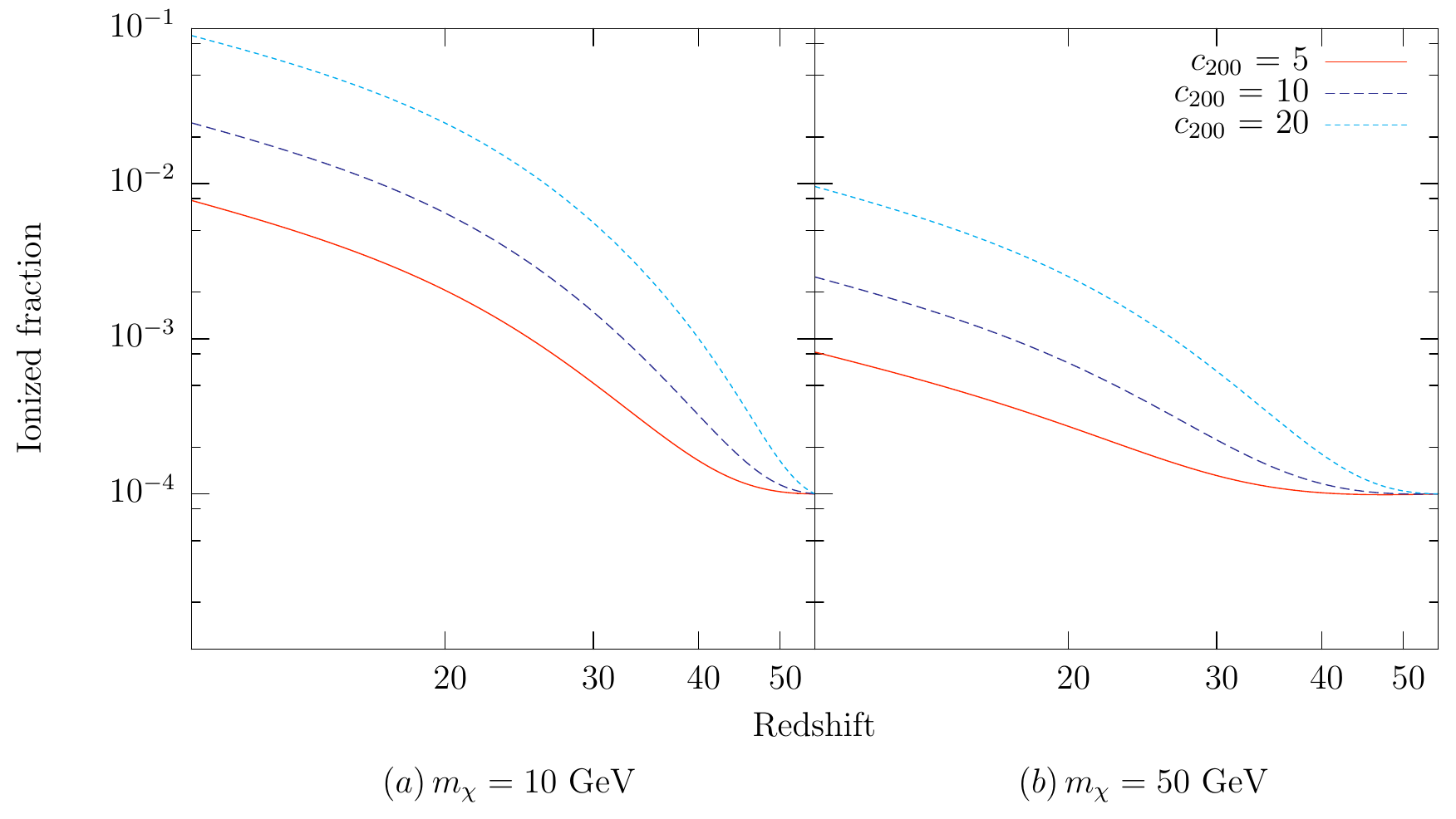}}
\end{center}
\caption{  Evolution of the ionized fraction $x_{\rm ion}$ with redshift, for different values of the concentration parameter $c_{\rm 200}$. (a) shows the case for $m_\chi = 10$ GeV while (b) is plotted for $m_\chi = 50$ GeV. The residual ionized fraction at a redshift $z=55$ was chosen to be $10^{-4}$. \label{fig3}  }
\end{figure}

\begin{figure}[!ht]
\begin{center}
\scalebox{0.8}{\includegraphics{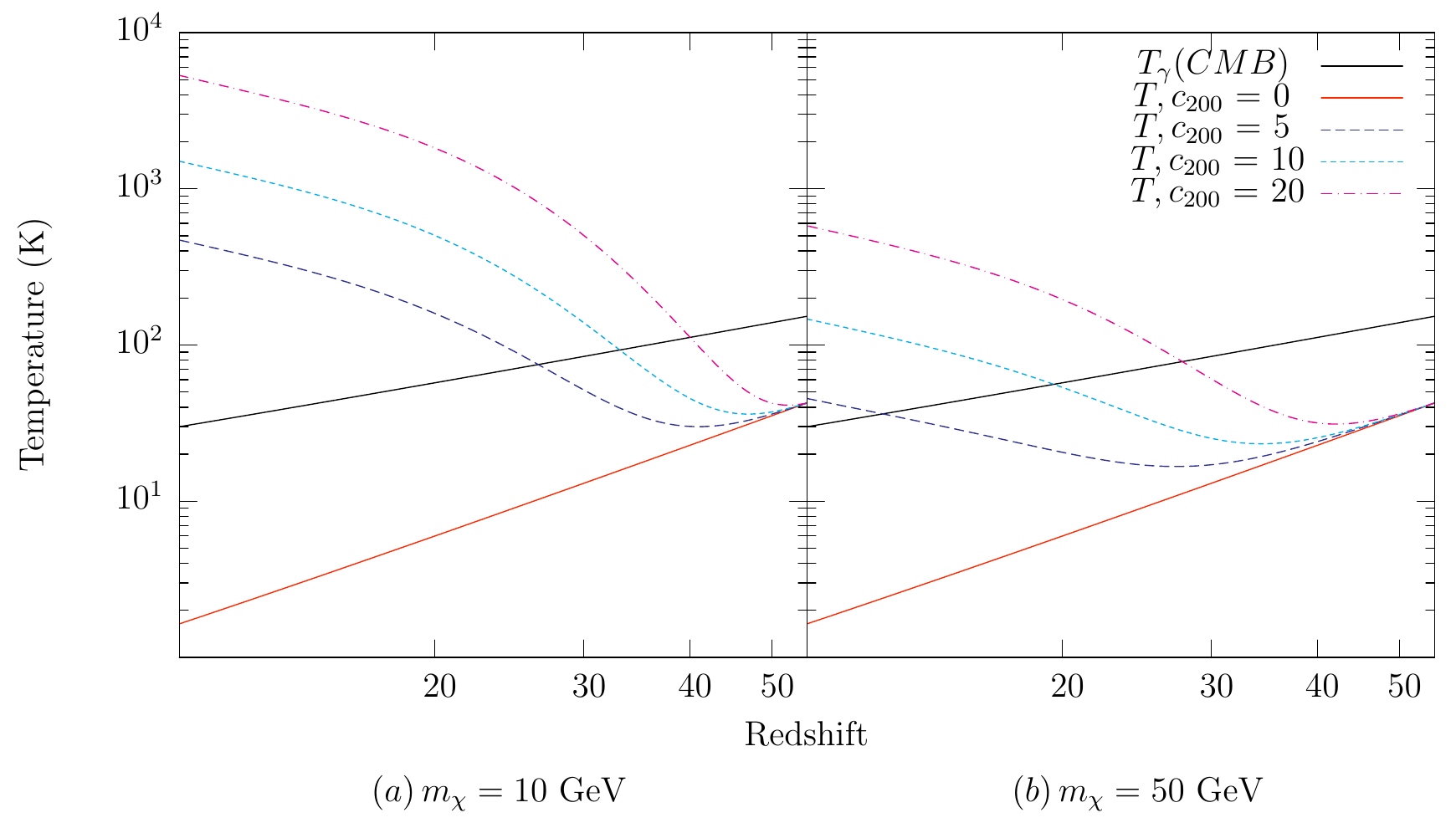}}
\end{center}
\caption{  Evolution of the gas temperature with redshift. (a) shows the case for $m_\chi = 10$ GeV, while (b) is plotted for $m_\chi = 50$ GeV. The solid red line shows the gas temperature $\sim (1+z)^2$ in the absence of dark matter heating. The solid black line shows the CMB temperature $\sim (1+z)$. The dashed lines show the evolution of the gas temperature for various concentration parameters. \label{fig4} }
\end{figure}

Figs. \ref{fig3} and \ref{fig4} show the evolution of the ionized fraction $x_{\rm ion}$ and the gas temperature $T$ with redshift. We note from Fig. \ref{fig3} that only models with small particle masses $m_\chi \sim 10$ GeV and concentration parameters $c_{\rm 200} \gtrsim 10$ are important from the point of view of reionization. The solid red line in Fig. \ref{fig4} shows the gas temperature $\sim (1+z)^2$ in the absence of dark matter heating. The solid black line shows the CMB temperature $\sim (1+z)$. The dashed lines show the gas temperature for different models with dark matter heating. Our results are in agreement with previous work \cite{chuzhoy}. In Fig. \ref{fig4}(a), the gas temperature eventually becomes larger than the CMB temperature for all the models considered. In Fig. \ref{fig4}(b), $T$ exceeds $T_\gamma$ only for the models with $c_{\rm 200}$ = 10 and 20.   

\section{Early reionization.}

Let us now consider different dark matter models and study the  epoch of reionization for a fixed optical depth. We then compare the expected CMB polarization power spectra for the different models.

\subsection{Optical depth and reionization redshift.}

The optical depth due to scattering of CMB photons with free electrons is given by
\beq
\tau(z) = \int_0^z \, c \, dz' \, \left( \frac{-dt}{dz'} \right ) \, \sigma_{\rm T} \, n_{\rm e}(z').
\label{tau}
\eeq
\begin{figure}[!ht]
\begin{center}
\scalebox{1}{\includegraphics{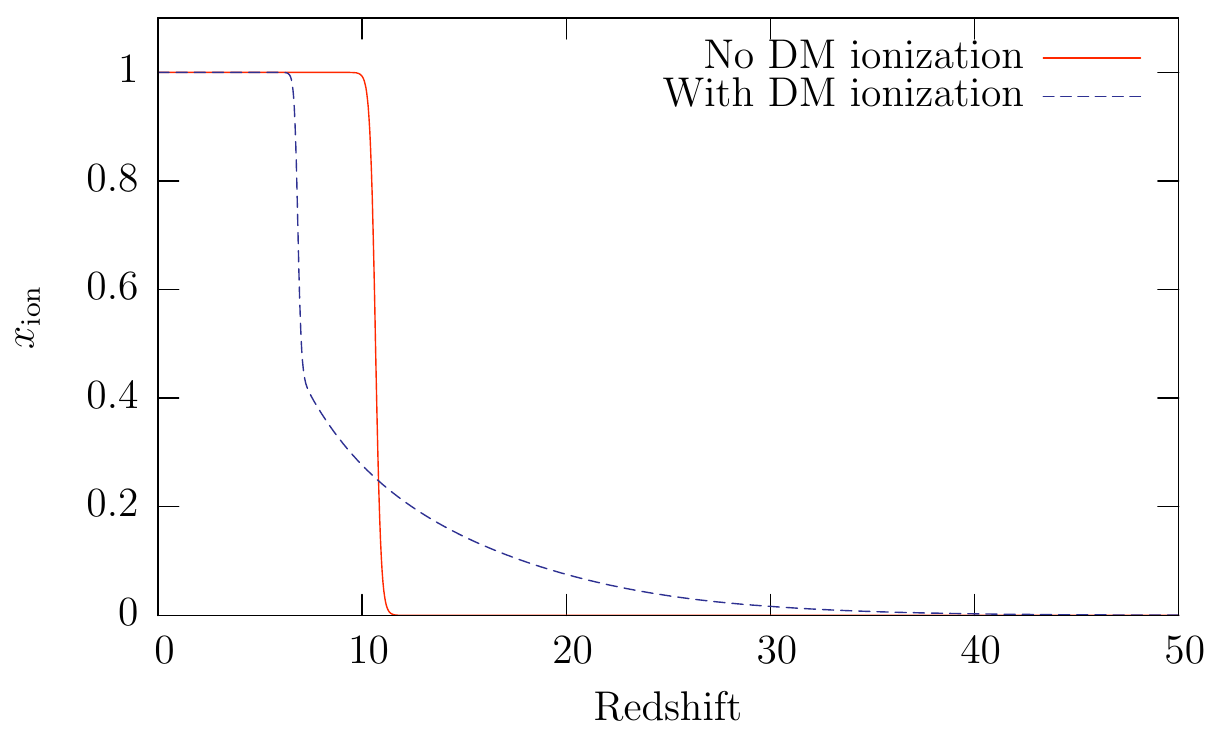}}
\end{center}
\caption{ Two different reionization histories. The solid (red) curve shows a sudden reionization scenario in which the Universe remains neutral until a certain redshift $z=z_\ast$, below which $x_{\rm ion} = 1$. The blue curve takes into account ionization by dark matter, for the model with $m_\chi = 2$ GeV and $c_{\rm 200} = 10$. $z_\ast$ is different from the previous case and $x_{\rm ion} \ne 0$ for $z > z_\ast$. Both curves give the same optical depth $\tau = 0.087$.  \label{fig5}  }
\end{figure}

Studies of quasar spectra have confirmed that the Universe is nearly completely ionized up to $z \approx 6$. Assuming H is ionized at $z=6$, He is singly ionized at $z=6$, and doubly ionized at $z=3$ \cite{He_ion}, we find $\tau(z=6) = 0.04$. As this value if less than the WMAP measured \cite{wmap} value of $\tau \approx 0.087$, we conclude that the Universe is at least partially ionized at redshifts $z > 6$. Let $z_\ast$ be the redshift below which $x_{\rm ion} = 1$. The simplest model of reionization is one in which $x_{\rm ion}(z) = 0$ for $z > z_\ast$. In this model, the Universe is assumed to be instantaneously reionized at $z=z_\ast$. Let us refer to this model as the ``sudden reionization'' scenario. While only an idealization, this model gives us a rough idea of the epoch of reionization.   Then, we can rewrite Eq. (\ref{tau}) as
\beq
\tau = 0.04 + \int_{6}^{z_\ast} \, c \, dz\, \left( \frac{-dt}{dz} \right ) \, \sigma_{\rm T}  \, n(z) + \int_{z_\ast}^{z_{\rm F}} \, c \, dz\, \left( \frac{-dt}{dz} \right ) \, \sigma_{\rm T}  \, n(z) x_{\rm ion} (z).
\eeq

\begin{table} {Variation of $z_{\ast}$ with $c_{\rm 200}$ and $m_\chi$.  \\ }

(a) $c_{\rm 200}$ = 10

\begin{tabular}  {|c| |c |c |c |c |c |c |c |c|} 
   \hline 
   $m_\chi$ (GeV) & 1 & 2 & 3 & 4 & 5 & 10 & 50 & $\infty$ \\  \hline  
   $z_\ast$ & $\cdots$ & 7.8 & 9.1 & 9.6 & 9.8 & 10.2 & 10.4 & 10.66  \\   
   \hline   
\end{tabular}

\vspace{0.2 in}

(b) $m_\chi$ = 2 GeV

\begin{tabular}  {|c| |c |c |c |c |c |c |c| } 
    \hline
     $c_{\rm 200}$ & 0 & 5 & 6 & 7 & 8 & 9 & 10\\  \hline  
     $z_\ast$ & 10.66 & 9.8 & 9.5 & 8.9 & 8.4 & 7.8 & 6.85 \\   
     \hline
\end{tabular}

\vspace{0.1in}

 \caption{ (a) shows how $z_\ast$ depends on the assumed particle mass $m_\chi$, with $c_{\rm 200}$ set equal to 10. (b) shows the variation of the reionization redshift $z_\ast$ with the assumed concentration parameter $c_{\rm 200}$ with the dark matter particle mass $m_\chi$ set equal to 2 GeV.  In each case, the total optical depth is 0.087. The cases $c_{\rm 200} = 0$ and $m_\chi = \infty$ correspond to no ionization by dark matter annihilation. }  
\end{table}
\begin{figure}[!h]
\begin{center}
\scalebox{0.8}{\includegraphics{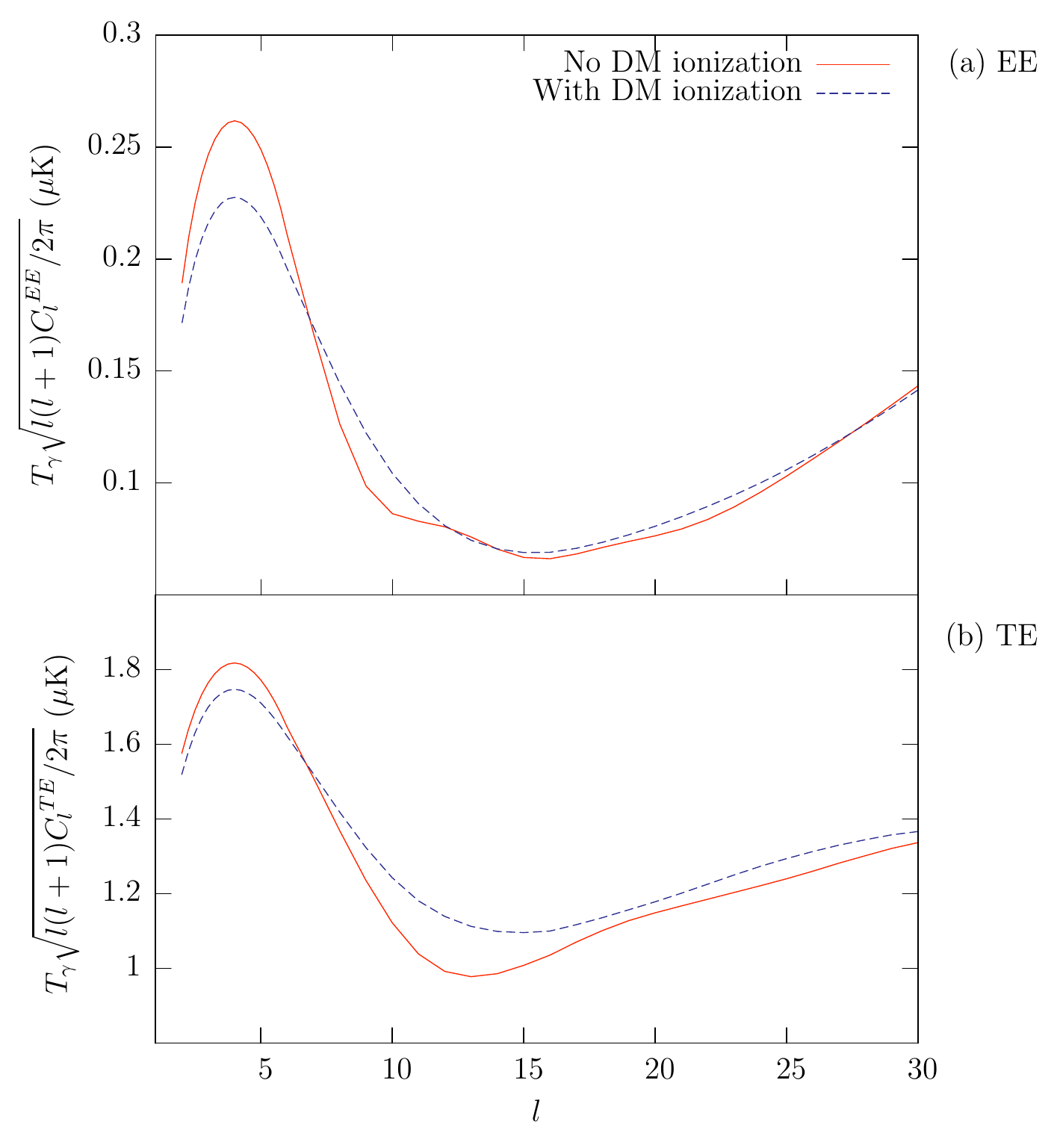}}
\end{center}
\caption{ CMB polarization power spectra. (a) shows the EE power spectrum, while (b) shows the TE power spectrum. The solid (red) curve is plotted for the sudden reionization scenario with no ionization by dark matter (the solid, red curve in Fig. 5). The dashed (blue) curve includes ionization by dark matter and corresponds to the dashed, blue curve of Fig. 5 ($m_\chi = 2$ GeV, $c_{\rm 200} = 10$). \label{fig6}}
\end{figure}

The sudden reionization scenario gives $z_\ast = 10.66$ assuming only ionized Hydrogen beyond $z=6$. When ionization by dark matter is taken into account, the value of $z_\ast$ is decreased as $x_{\rm ion} \ne 0$ for $z > z_\ast$.  Fig. 5 shows 2 different reionization histories, giving the same value of $\tau = 0.087$. The solid (red) curve shows the sudden reionization scenario in which $x_{\rm ion} = 0$ for $z > z_\ast = 10.66$ and $x_{\rm ion} = 1$ for $z < z_\ast$. The dashed (blue) curve shows an alternate model which includes dark matter ionization for $z > z_\ast = 6.85$. $x_{\rm ion} = 1$ for $z < z_\ast$. $m_\chi = 2$ GeV and $c_{\rm 200} = 10$.

 Table 1 shows $z_\ast$ for different values of  $m_\chi$ and $c_{\rm 200}$. $z_\ast$ is considerably lowered only for very light dark matter particles ($\sim$ 1 GeV), and for large concentration parameters ($c_{\rm 200} \sim 10$).

\subsection{ Effect on CMB polarization spectra.}

The partial ionization of the Universe at high redshifts increases the scattering of CMB photons with free electrons, thus influencing CMB polarization. The increased scattering at high redshifts results in a transfer of power from low multipoles to higher multipoles in the large angle EE and TE polarization spectra, compared to the sudden reionization scenario. Fig. \ref{fig6} shows the EE and TE polarization power spectra computed using the CAMB software \cite{camb}, for the two reionization histories considered in Fig. \ref{fig5}. The solid (red) line corresponds to the sudden reionization scenario in which $x_{\rm ion} = 1$ for $z < z_\ast$ and $0$ for $z > z_\ast$. The dashed (blue) line corresponds to the reionization history with DM annihilation taken into account, for a model with $m_\chi = 2$ GeV, $c_{\rm 200} = 10$. $x_{\rm ion} = 1$ for $z < z_\ast$.  (compare with Fig. \ref{fig5}). The scattering of electrons at high redshifts results in a reduced amplitude for the first peak at $l \approx 4$ and greater power at slightly higher $l$. For $l > 30$, there is very little difference between the spectra. It may be possible to distinguish these curves with future experiments such as Planck and CMBPol \cite{pol,haiman}. However, differentiating between the dark matter models and a model of gradual reionization by astrophysical sources (as opposed to the sudden reionization model) will be considerably more challenging. It was recently shown \cite{hooper} that in certain dark matter models with a large value of $\langle \sigma_{\rm a} v \rangle$, ionization by dark matter annihilation is substantially larger. This would result in large modifications to the CMB polarization power spectrum as well.

\section{The Hydrogen 21 cm line.}

Hydrogen 21 cm cosmology is an important tool to study reionization and the early Universe \cite{Hydrogen}. The 21 cm differential brightness temperature (brightness temperature relative to the CMB) is given by (see for e.g., \cite{furl1,furl_review}):
\beq
T_{\rm b}(z)   \approx 27 \, \textrm{mK} \;\; \sqrt{ \frac{1+z}{10} } \; (1-x_{\rm ion}) \, \frac{n}{n_{\rm 0}} \, \left( 1 - \frac{T_\gamma}{T_{\rm s}} \right ) \left[ \frac{ H(z)/(1+z) }{dv_{||} / dr_{||} } \right ].
\label{dTb}
\eeq
$n_{\rm 0}$ is the spatial average of $n$. $T_{\rm s}$ is the spin temperature given by
\beq
T^{-1}_{\rm s} \approx \frac{    T^{-1}_\gamma + \left( \xi_{\rm c} + \xi_{\alpha} \right ) T^{-1} }{1 + \xi_{\rm c} + \xi_{\alpha} }.
\label{Ts}
\eeq
$\xi_{\rm c}(z,x_{\rm ion},T)$ is called the collisional coupling coefficient \cite{furl_review}:
\beq
\xi_{\rm c} = \frac{n(1+z)^3}{A_{\rm 10}} \frac{T_\ast}{T_{\gamma,0} (1+z)} \, \left[ x_{\rm ion} \kappa^{\rm e} + (1-x_{\rm ion}) \kappa^{\rm H} \right ].
\label{f}
\eeq
$A_{\rm 10}$ is the Einstein coefficient for spontaneous emission. $T_\ast = E_{\rm 10}/k_{\rm b}$, with $E_{\rm 10}$ being the energy difference between the singlet and triplet levels. $\kappa^{\rm e}(T)$ and $\kappa^{\rm H}(T)$ are rate coefficients for collisions with electrons and Hydrogen atoms respectively. At low temperatures, the ionization fraction is very low, and we expect $\kappa^{\rm H}$ to be the dominant term. $\kappa^{\rm e}$ becomes relevant for larger values of $x_{\rm ion}$ which are associated with higher $T$. We may neglect collisions with protons for large $T$ \cite{furl_furl}.  $dv_{||}/dr_{||}$ is the gradient of the proper velocity along the line of sight. $\xi_\alpha$ is the Wouthuysen-Field (Ly-$\alpha$) coupling term \cite{furl_review}:
 \beq
 \xi_\alpha = \frac{16 \pi^2 c f_\alpha}{27 A_{\rm 10}} \, \frac{T_\ast}{T_{\gamma,0}(1+z)} \, \left( \frac{e^2}{mc^2} \right ) \int d\nu J(\nu) \Phi(\nu)
 \label{xi_alpha_1}
 \eeq

\begin{figure}[!ht]
\begin{center}
\scalebox{0.7}{\includegraphics{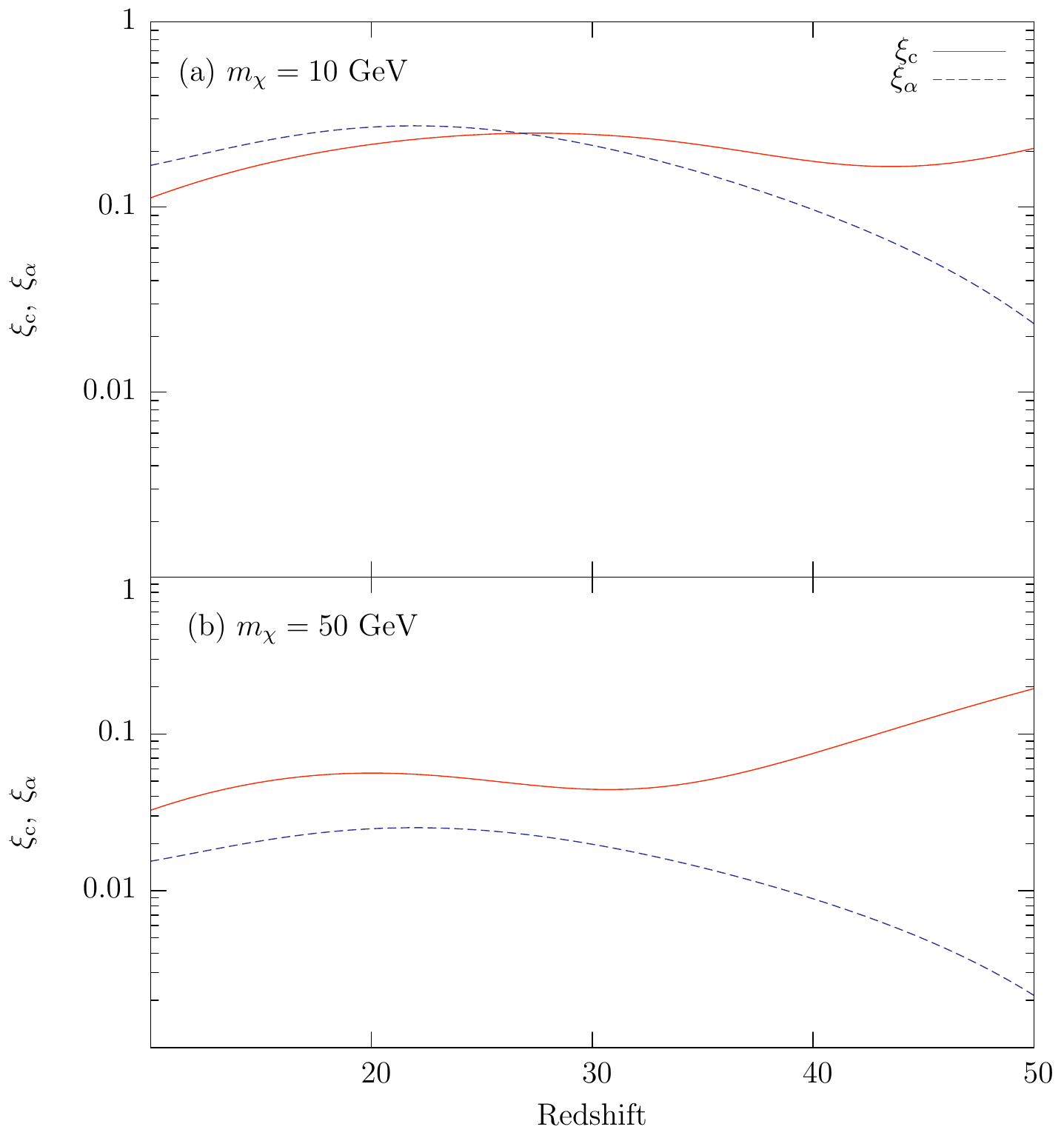}}
\end{center}
\caption{  Collisional ($\xi_{\rm c}$) and Ly-$\alpha$ ($\xi_\alpha$) coupling terms for particle masses $m_\chi = 10$ and $50$ GeV. $c_{\rm 200} = 10$. \label{fig7} }
\end{figure}

Fig. \ref{fig7} shows $\xi_{\rm c}$ and $\xi_\alpha$ for $m_\chi$ = 10 and 50 GeV, with $c_{\rm 200} = 10$. For redshifts $z < 25$, both coupling coefficients are important for light dark matter particles $m_\chi \sim 10$ GeV. For heavier dark matter particles $m_\chi \sim 50$ GeV, $\xi_{\rm c} \gg \xi_\alpha$. $f_\alpha = 0.4162$ is the oscillator strength of the transition.  $J(\nu)$ is the number of Ly-$\alpha$ photons per unit area, per time, per frequency, and per solid angle. Following \cite{furl1}, we make the assumption that one half of the energy that goes into collisional excitations results in the generation of Ly-$\alpha$ photons. With the simple ansatz that a third of the total energy absorbed goes into collisional excitations, we have a fraction $\eta_\alpha = 1/6$ of the total energy resulting in Ly-$\alpha$ photons.  $\Phi(\nu)$ is the line profile satisfying $\int d\nu \, \Phi(\nu) = 1$. We will assume that $\Phi(\nu)$ is sharply peaked about $\nu = \nu_\alpha \approx 2.5 \times 10^{15}$ Hz. $J$ may be expressed as \cite{furl1}:
\beq
J \approx  \frac{\eta_\alpha c}{4 \pi} \, \frac{1}{H_{\rm 0} \sqrt{\Omega_{\rm m}} (1+z)^{3/2} } \frac{E}{h \nu_\alpha} \frac{1}{\nu_\alpha}
\eeq
$E(z,m_\chi,c_{\rm 200})$ is given by Eq. \ref{E_equation}. Using Eq. \ref{xi_alpha_1} with $\Omega_{\rm m} = 0.258, h = 0.71$, we find
\beq
\xi_\alpha \approx 0.012 \left( \frac{21}{1+z} \right )^{5/2} \, \left( \frac{E}{10^{-20} \, {\rm eV cm}^{-3} {\rm s}^{-1} } \right )
\eeq

Using Eq. (\ref{Ts}), we can rewrite Eq. (\ref{dTb}) as:
\beq
T_{\rm b} = 27 \, \textrm{mK} \;\; \sqrt{ \frac{1+z}{10} } \; (1-x_{\rm ion}) \, \frac{n}{n_{\rm 0}} \, \frac{\xi}{1+\xi} \, \left( 1 - \frac{T_\gamma}{T} \right ) \left[ \frac{ H(z)/(1+z) }{dv_{||} / dr_{||} } \right ].
\label{diff_T}
\eeq
where $\xi = \xi_{\rm c} + \xi_\alpha$.

\begin{figure}[!ht]
\begin{center}
\scalebox{0.8}{\includegraphics{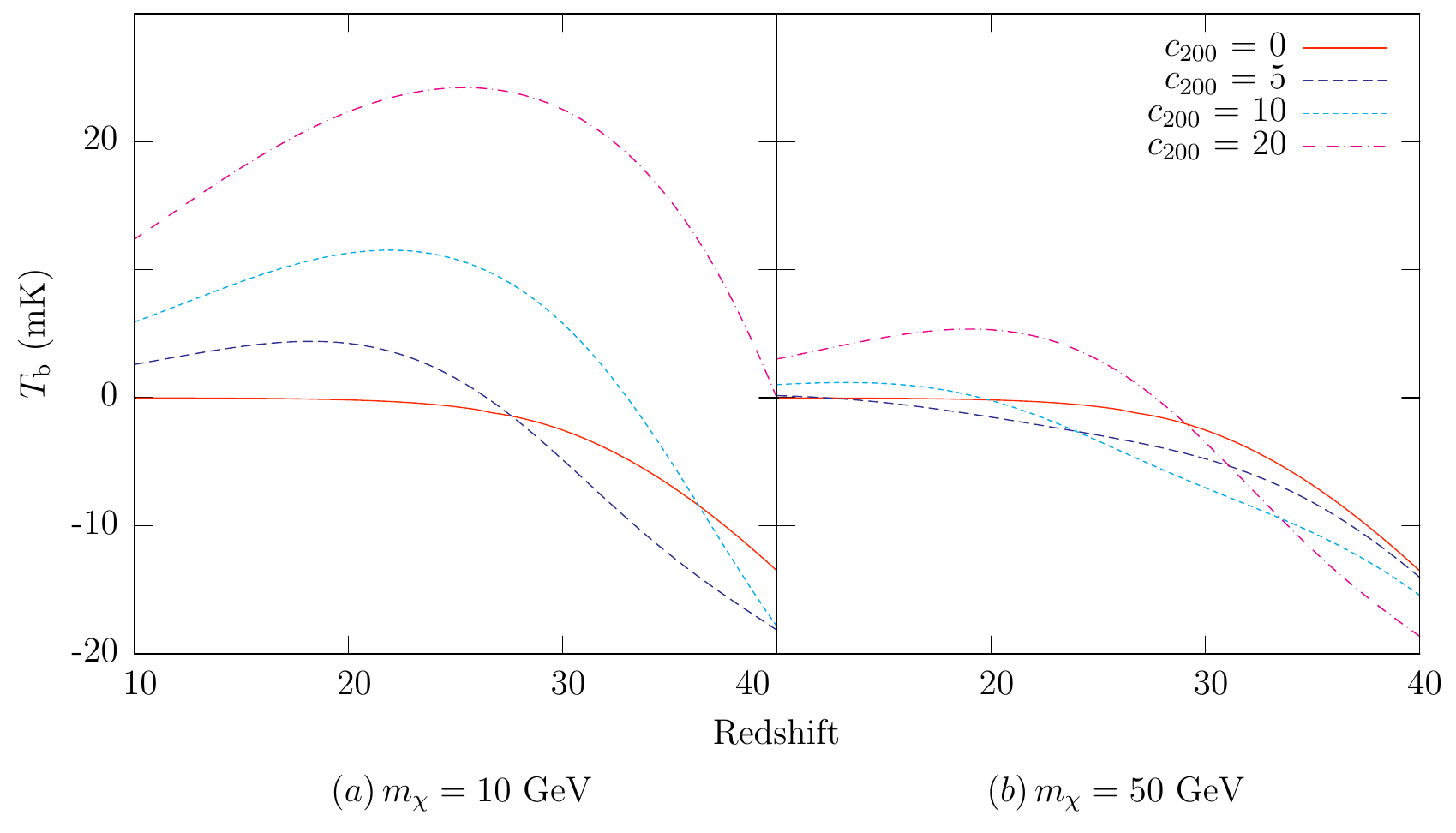}}
\end{center}
\caption{  H21 cm differential brightness temperature $T_{\rm b}$. The solid (red) line shows $T_{\rm b}$ with no dark matter heating, while the dashed lines are drawn for models with $c_{\rm 200} = 5,10$, and 20. Shown are the plots for $(a) m_\chi = 10$ GeV and $(b) m_\chi = 50$ GeV. \label{fig8} }
\end{figure}

Fig. \ref{fig8} shows the spatially averaged value of $T_{\rm b}$ for different values of the concentration parameter $c_{\rm 200}$, for masses (a) $m_\chi$ = 10  and (b) 50 GeV. The solid (red) curve shows $T_{\rm b}$ in the absence of dark matter heating ($c_{\rm 200} = 0$). Let us refer to this temperature as $T_{\rm b}$(no DM). The gas temperature with no dark matter heating is labeled $T$(no DM).  For some values of $c_{\rm 200}$, the differential brightness temperature $T_{\rm b}$ starts out smaller than $T_{\rm b}$(no DM), but soon crosses this curve. This is caused by the gas temperature $T$ being smaller than the CMB temperature $T_\gamma$, yet larger than $T$(no DM). Eventually $T$ becomes larger than  $T_\gamma$ (compare with Fig. \ref{fig4}). For $(c_{\rm 200} = 20, m_\chi$ = 10 GeV), $T_{\rm b}$ is larger than $T_{\rm b}$(no DM) for all $z$. Conversely, for $(c_{\rm 200} = 5, m_\chi$ = 50 GeV), $T_{\rm b}$ is smaller than $T_{\rm b}$(no DM) for all $z$. We will explore the redshift variation of $T_{\rm b}$ further when we discuss the power spectrum.

\subsection{Perturbations in the brightness temperature.}
Let us now consider perturbations in the differential brightness temperature $T_{\rm b}$. Eq. (\ref{diff_T}) is sensitive to fluctuations in the baryon density, the ionized fraction, and the gas temperature. Let us perturb $x_{\rm ion}$, $n$, and $T$, and expand to linear order:
\bea
1-x_{\rm ion} &=& 1 - x_{\rm ion,0} -x_{\rm ion,0} \delta_{\rm x} \n
n &=& n_{\rm 0} + n_{\rm 0} \delta_{\rm n} \n
1 - \frac{T_\gamma}{T} &=& 1 - \frac{T_\gamma}{T_{\rm 0}} + \frac{T_\gamma}{T_{\rm 0}} \delta_{\rm T} \n
\xi &=& \xi_{\rm 0} + \left[ \frac{\partial \xi_{\rm 0}}{\partial n} \, n_{\rm 0} \delta_{\rm n} + \frac{\partial \xi_{\rm 0}}{\partial T} \, T_{\rm 0} \delta_{\rm T} + \frac{\partial \xi_{\rm 0}}{\partial x_{\rm ion}} \, x_{\rm ion,0} \delta_{\rm x} \right ],
\eea
where $x_{\rm ion,0}, n_{\rm 0}$, and $T_{\rm 0}$ are spatially averaged values, and $\delta_{\rm x}, \delta_{\rm n}$, and $\delta_{\rm T}$ are fractional perturbations in the ionized fraction, baryon density, and gas temperature respectively.  $\xi_{\rm 0}$ is evaluated at $(x_{\rm ion,0}, n_{\rm 0}, T_{\rm 0})$.

The fractional perturbation in the brightness temperature in Fourier space is then found to be
\beq
\delta_{\rm 21}(z, \vec k) = \delta_{\rm n}(z, \vec k) \left[ \beta_{\rm n}(z) + \left( \hat n \cdot \hat k \right )^2 \right ]  + \delta_{\rm T}(z, \vec k) \beta_{\rm T}(z) + \delta_{\rm x}(z, \vec k) \beta_{\rm x}(z),
\label{delta_21}
\eeq
where $\hat n$ is the direction of the line of sight, $\vec k$ is the Fourier wave vector, and
\bea
\beta_{\rm n} &=&  \left[ 1 + \frac{1}{1+\xi_{\rm 0}} \frac{\partial \ln \xi_{\rm 0}}{\partial \ln n}  \right ] \n
\beta_{\rm T} &=&  \frac{T_\gamma/T_{\rm 0}}{1 - T_\gamma/T_{\rm 0}} +  \frac{1}{1+\xi_{\rm 0}}  \frac{\partial \ln \xi_{\rm 0}}{\partial \ln T}  \n
\beta_{\rm x} &=&   \left[ -\frac{x_{\rm ion,0}}{1-x_{\rm ion,0}} + \frac{1}{1+\xi_{\rm 0}}  \frac{\partial \ln \xi_{\rm 0}}{\partial \ln x_{\rm ion}}  \right ].
\eea
To compute the derivates of $\xi = \xi_{\rm c} + \xi_\alpha$, let us first consider perturbations in the density of dark matter halos. The energy absorbed by the gas at any redshift $z$ depends on the halo density for redshifts $z' > z$, as seen in Eq. (\ref{E_equation}). The integration over redshifts in Eq. (\ref{E_equation}) results in the perturbations averaging out as statistically, overdense regions are as likely to exist as underdense regions. As a first approximation, we may ignore perturbations in halo density. The fractional perturbation caused by the dark matter halos $\propto 1/\sqrt{N}$ where $N$ is the number of halos in a mean free path volume. Hence this assumption is valid provided $\sqrt{N} \gg 1$.  The mean free path of a high energy photon at redshift $z \approx \left[ n (1+z)^3 \sigma \right ]^{-1} \approx 100$ Mpc for $z=25$, where $n$ is the comoving baryon number density, and we assumed $\sigma = \sigma_{\rm T}$ (although for high energy photons, $\sigma < \sigma_{\rm T}$, which makes the mean free path larger). The number density of halos at redshift $z$ is
\beq
 n_{\rm halo}(z) = (1+z)^3 \int_{M_{\rm min}}^\infty dM \, \frac{dn_{\rm h}}{dM} 
 \eeq
$\approx 10^{18} \, \textrm{Mpc}^{-3} ( 10^{12} \, \textrm{Mpc}^{-3} )$ for $M_{\rm min} = 10^{-6} M_\odot ( 1 \, M_\odot )$, at $z=25$. The corresponding mass density $\approx 8 \times 10^{13} M_\odot$ Mpc$^{-3} ( 10^{13} \, M_\odot$ Mpc$^{-3} )$, about 15\% (2\%) of the dark matter density at $z=25$. $dn_{\rm h}/dM$ is given by Eq. [\ref{press_schechter}]. Thus $\sqrt{N} \approx 10^9 - 10^{12}$ depending on $M_{\rm min}$.  Using Eq. (\ref{f}) and $\xi_\alpha \propto E \propto n$, we have:
\bea
\frac{\partial \ln \xi}{\partial \ln n} &=& 1 \n
\frac{\partial \ln \xi}{\partial \ln T} &=& \frac{ T\left[ x_{\rm ion} \, \partial\kappa^{\rm e}/\partial T + (1-x_{\rm ion}) \partial\kappa^{\rm H} / \partial T \right ] }{ x_{\rm ion} \kappa^{\rm e} + (1-x_{\rm ion}) \kappa^{\rm H} }\n
\frac{\partial \ln \xi}{\partial \ln x_{\rm ion}} &=& \frac{ x_{\rm ion} \left[ \kappa^{\rm e} - \kappa_{\rm H} \right ] } { x_{\rm ion} \kappa^{\rm e} + (1-x_{\rm ion}) \kappa^{\rm H} }.
\eea
We show that for redshifts $z < 30$, $\beta_{\rm n}$ is often the dominant term.

\subsection{The multi-frequency angular power spectrum.}

Let us expand $\delta_{21}$ as a sum over spherical harmonics:
\beq
\delta_{\rm 21} (z, \hat n) = \sum_{l,m} a_{lm}(z) \, Y_{lm} (\hat n),
\eeq
where
\beq
a_{lm}(z) = \int d\Omega \, Y^\ast_{lm}(\hat n) \; \int \frac{d^3 k}{2\pi^3} \left[ \exp -ikr(\hat k \cdot \hat n) \right ] \delta _{\rm 21} (z,\vec k),
\eeq
and $r$ is the comoving distance
\beq
r(z) = r[\nu(z)] = \int_0^z dz' \frac{c}{H(z')} \approx \frac{2c}{H_{\rm 0}\sqrt{\Omega_{\rm m}}} \left[ 1 - \left( 1+z \right )^{-1/2} \right ],
  \label{comov}
\eeq
where we ignored $\Omega_\Lambda$. $\nu(z)$ is the redshifted 21 cm line frequency
\beq
\nu(z) = \frac{\nu_{\rm 0}}{1+z},
\eeq
where $\nu_{\rm 0} = c / 21.1$cm = 1.42 GHz. Setting $x = kr$, and using Eq. (\ref{delta_21}) and the identities
\bea
\int d\Omega \,   Y^\ast_{lm} (\hat n)  e^{ -ix \hat k \cdot \hat n} &=& 4 \pi (-i)^l j_l(x) Y^\ast_{lm}(\hat k) \n
\int d\Omega \,   Y^\ast_{lm} (\hat n) e^{-ix \hat k \cdot \hat n} (\hat k \cdot \hat n)^2 &=& -4 \pi (-i)^l \frac{\partial^2 j_l(x)}{\partial x^2} Y^\ast_{lm}(\hat k),
\eea
we find the following expression for $a_{lm}$:
\beq
a_{lm}(\nu) = 4 \pi (-i)^l \, \int \frac{d^3k}{(2\pi)^3} Y^\ast_{lm}(\hat k) \, \left[ \delta_{\rm n} \left \{ \beta_{\rm n} j_l(x) - \frac{ \partial^2 j_l(x) }{\partial x^2} \right \} + \delta_{\rm T} \beta_{\rm T} j_l(x) + \delta_{\rm x} \beta_{\rm x} j_l(x) \right ]
\label{a_lm}
\eeq
We can now construct the variance $C_l$ defined as
  \beq
  C_l (\nu, \Delta\nu) \, \delta_{ll'} \, \delta_{mm'} = \langle a_{lm} (\nu) \, a^\ast_{l'm'} (\nu') \rangle = 16 \pi^2 (-i)^l i^{l'} \int \frac{d^3 k}{2 \pi^3} \, \frac{d^3 k'}{2 \pi^3} \; Y^\ast_{lm} (\hat k) \, Y_{l' m'} (\hat k')  \; \mathcal{P}(z,z',\vec k, \vec k') 
  \label{Cl_nu_nu'}
  \eeq
$\mathcal{P}$ is given by
    \bea
  \mathcal{P} &=& \langle \delta_{\rm n}(z) \delta^\ast_{\rm n} (z') \rangle \left[ \beta_{\rm n}(z) j_l (x) - \partial ^2 j_l(x) / \partial x^ 2 \right ] \, \left[ \beta_{\rm n}(z') j_{l'} (x') - \partial ^2 j_{l'}(x') / \partial x'^ 2 \right ] \n
        &+& j_l (x) \left[ \beta_{\rm n}(z') j_{l'} (x') - \partial ^2 j_{l'}(x') / \partial x'^ 2 \right ] \, \left[ \beta_{\rm T} (z) \langle \delta_{\rm T} (z) \delta^\ast_{\rm n}(z') \rangle + \beta_{\rm x} (z) \langle \delta_{\rm x} (z) \delta^\ast_{\rm n}(z') \rangle \right ] \n
   &+& j_{l'} (x') \left[ \beta_{\rm n}(z) j_l (x) - \partial ^2 j_l(x) / \partial x^ 2 \right ] \, \left[ \beta_{\rm T} (z') \langle \delta_{\rm n} (z) \delta^\ast_{\rm T}(z') \rangle + \beta_{\rm x} (z') \langle \delta_{\rm n} (z) \delta^\ast_{\rm x}(z') \rangle \right ] \n
      &+& j_l(x) j_{l'}(x') \left[ \beta_{\rm T}(z) \beta_{\rm T}(z') \langle \delta_{\rm T} (z) \delta^\ast_{\rm T}(z') \rangle + \beta_{\rm x}(z) \beta_{\rm x}(z') \langle \delta_{\rm x} (z) \delta^\ast_{\rm x}(z')   \rangle  \right. \n 
        & & \hspace{0.65in}  \left. + \, \beta_{\rm T}(z) \beta_{\rm x}(z') \langle \delta_{\rm T} (z) \delta^\ast_{\rm x}(z') \rangle   + \, \beta_{\rm x}(z) \beta_{\rm T}(z') \langle \delta_{\rm x} (z) \delta^\ast_{\rm T}(z') \rangle  \right ].
\eea
$x' = k'r'$. Let us now make the approximation that $| \nu' - \nu | = \Delta \nu \ll \nu_0$:
\beq
|r' - r| = \Delta r \approx \frac{c}{H_{\rm 0} \sqrt{\Omega_{\rm m}}} \sqrt{1+z} \, \left | \frac{ \Delta \nu }{\nu_{\rm 0}} \right | 
\eeq
For small values of $| \Delta \nu / \nu_{\rm 0} |$, we may make the following approximations:
\bea
\beta_i (z) &\approx& \beta_i (z') \n
\langle \delta_i (z, \vec k) \delta^\ast_j (z', \vec k') \rangle &\approx& \langle \delta_i (z, \vec k) \delta^\ast_j (z, \vec k') \rangle \n
     &=&  (2 \pi)^3 \delta^3 (\vec k - \vec k') P_{ij}(z, k) = (2 \pi)^3 \delta^3 (\vec k - \vec k') P_{ji}(z, k)
\eea
where $i$ and $j$ may stand for the baryon number density, ionized fraction, or gas temperature fluctuations. Using the above approximations, setting $x=kr, x'=kr'$, and the identity
\beq
\int d\Omega \,   Y_{lm} (\hat n) \; Y^\ast_{l'm'} (\hat n)  = \delta_{ll'} \delta_{mm'}
\eeq 
we may simplify Eq. (\ref{Cl_nu_nu'}):
\bea
C_l(\nu,\Delta\nu) &=& \frac{2}{\pi} \int_0^\infty dk \; k^2 \times \n
& & \left[ \;\;\;\; P_{\rm nn}(z,k) \left \{ \beta_{\rm n}(z) j_l(x) - \partial^2 j_l(x) / \partial x^2 \right \} \left \{ \beta_{\rm n}(z) j_l(x') - \partial^2 j_l(x') / \partial x'^2 \right \} \right . \n
&+& \left \{ \beta_{\rm T}(z) P_{\rm nT}(z,k) + \beta_{\rm x}(z) P_{\rm nx}(z,k) \right \} \left [ j_l(x) \left \{ \beta_{\rm n}(z) j_l(x') - \partial^2 j_l(x') / \partial x'^2 \right \}  \right . \n
& & \left. \hspace{2.3 in} + \, j_l(x') \left \{ \beta_{\rm n}(z) j_l(x) - \partial^2 j_l(x) / \partial x^2 \right \} \right ] \n
&+& \left. j_l(x) j_l(x') \left \{ \beta^2_{\rm T} P_{\rm TT}(z,k) + 2 \beta_{\rm T} \beta_{\rm x} P_{\rm Tx}(z,k) + \beta^2_{\rm x} P_{\rm xx}(z,k) \right \} \hspace{0.2in} \right ] 
\label{power_spectrum}
\eea

\subsection{Fluctuations in $x_{\rm ion}$ and $T$.}

 To compute $C_l$, we first  need to compute the various power spectra in Eq. (\ref{power_spectrum}). Since we are ignoring perturbations in halo density, the fractional perturbation in $E$ is then $\delta_{\rm E} = \delta_{\rm n}$. Let us express the fractional perturbations in the ionized fraction $(\delta_{\rm x})$ and the gas temperature $(\delta_{\rm T})$ at redshift $z$ as:
\bea
  \delta_{\rm x}(z, \vec k) &=& S_{\rm x}(z) \, \delta_{\rm n}(z, \vec k) \n
  \delta_{\rm T}(z, \vec k) &=& S_{\rm T}(z) \, \delta_{\rm n}(z, \vec k),
  \label{delta_approx}
\eea
i.e. we make the approximation that the $k$ dependent part of $\delta_{\rm x}$  and $\delta_{\rm T}$ follows the form of $\delta_{\rm n}$. This is motivated by the dependence of Eq. (\ref{E_equation}), on the baryon density. $S_{\rm x}$ and $S_{\rm T}$ give us the time evolution of the perturbations. With this approximation, we consider perturbations in Eq. (\ref{ion}) and Eq. (\ref{T}), to obtain the linearized equations:
\bea
\frac{d S_{\rm x}}{dz} &=& \left[ \frac{n \, x_{\rm ion} \alpha \sqrt{1+z}}{H_{\rm 0} \sqrt{\Omega_{\rm m}}}  \right ] \left (1 + 2S_{\rm x} + S_{\rm T} \frac{\partial \ln \alpha}{\partial \ln T} \right ) + S_{\rm x} \left[ \frac{1}{1+z} - \frac{1}{x_{\rm ion}} \frac{dx_{\rm ion}}{dz} +  \frac{ \mu \, \eta_{\rm ion} \,  E}{n H_{\rm 0} \sqrt{\Omega_{\rm m}}(1+z)^{11/2} }  \right ] \n
\frac{d S_{\rm T}}{dz} &=& \frac{1}{H_{\rm 0} \sqrt{\Omega_{\rm m}}\left(1+f_{\rm He} \right )}  \; \frac{x_{\rm ion}}{ t_{\rm c} (1+z)^{5/2}} \left[ S_{\rm x} \left( 1 - \frac{T_\gamma}{T} \right ) + S_{\rm T} \right ] + S_{\rm T} \left[ \frac{3}{1+z} - \frac{1}{T} \frac{dT}{dz} \right ],
\eea
where we used $x_{\rm ion} \ll 1$. The power spectra then become
\bea
P_{\rm TT}(z,k) &=& S^2_{\rm T}(z) P_{\rm nn}(z,k) \n
P_{\rm xx}(z,k) &=& S^2_{\rm x}(z) P_{\rm nn}(z,k) \n
P_{\rm nT}(z,k) &=& S_{\rm T}(z) P_{\rm nn}(z,k) \n
P_{\rm nx}(z,k) &=& S_{\rm x}(z) P_{\rm nn}(z,k) \n
P_{\rm Tx}(z,k) &=& S_{\rm T}(z) S_{\rm x}(z) P_{\rm nn}(z,k).
\eea
Eq. (\ref{power_spectrum}) can be simplified to the form
\bea
C_l (\nu, \Delta\nu) &=& \frac{2}{\pi} \, \frac{1}{\left( 1+z \right )^2} \; \int_0^\infty dk \, k^2 P(k) \times  \\
& & \left[ \beta^2(z) j_l(x) j_l(x') - \beta(z) \left \{ j_l(x) \frac{\partial^2 j_l(x')}{\partial x'^2} + j_l(x') \frac{\partial^2 j_l(x)}{\partial x^2} \right \} + \frac{\partial^2 j_l(x)}{\partial x^2} \frac{\partial^2 j_l(x')}{\partial x'^2} \right ] \nonumber
\eea
where $\beta(z) = \beta_{\rm n}(z) + S_{\rm T}(z) \beta_{\rm T}(z) + S_{\rm x}(z) \beta_{\rm x}(z)$, and we assumed $P_{\rm nn}(z,k) = P(z,k) = P(k)/(1+z)^2$. 

\begin{figure}[!ht]
\begin{center}
\scalebox{0.8}{\includegraphics{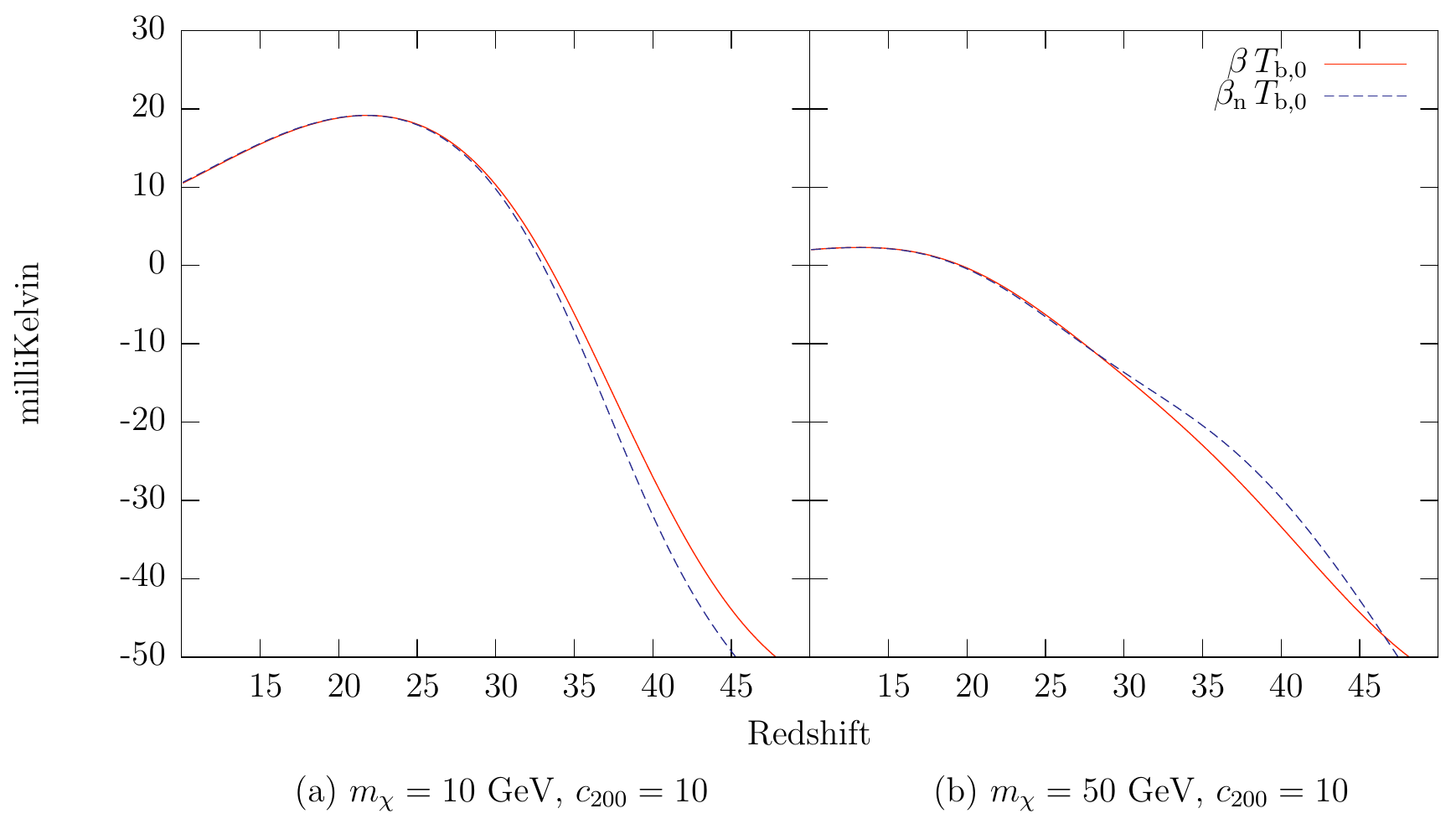}}
\end{center}
\caption{  $\beta T_{\rm b,0}$ and $\beta_{\rm n} T_{\rm b,0}$ as a function of redshift $z$. The curves are nearly identical at low redshifts which implies that our results are not very sensitive to the form of $\delta_{\rm T}$ and $\delta_{\rm x}$.  \label{fig9} }
\end{figure}

\begin{figure}[!ht]
\begin{center}
\scalebox{0.8}{\includegraphics{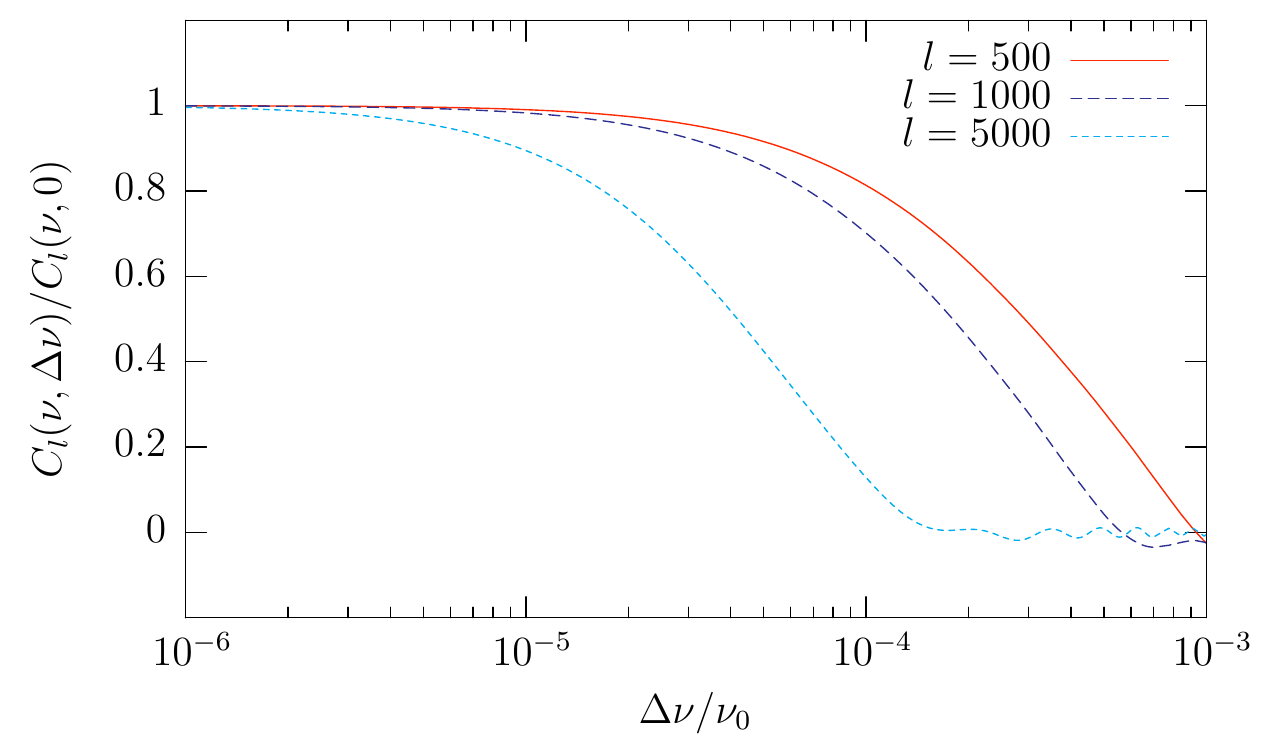}}
\end{center}
\caption{  Variation of $C_l(\nu, \Delta\nu)$ with $\Delta\nu / \nu_{\rm 0}$, at $z=25 \, (\nu(z) \approx$ 55 MHz). The decrease in power with $\Delta\nu$ may be used to separate the Hydrogen 21 cm signal from the large foregrounds.  \label{fig10} }
\end{figure}

Fig. \ref{fig9} shows $\beta T_{\rm b,0}$ and $\beta_{\rm n} T_{\rm b,0}$ for particle masses $m_\chi$ = 10 and 50 GeV, with initial conditions $S_{\rm T} = S_{\rm x} = 1$ at $z=55$. $T_{\rm b,0}$ is the spatially averaged value of $T_{\rm b}$. The difference between the two curves is a measure of the importance of fluctuations in the ionized fraction and temperature. In the redshift range $10<z<30$, the two curves are nearly identical, and thus our results are not very sensitive to fluctuations in the ionized fraction and temperature in this range, except when $T \approx T_\gamma$. 

Fig. \ref{fig10} shows the normalized multifrequency power spectrum $C_l (\nu, \Delta\nu)$ as a function of $\Delta\nu/\nu_{\rm 0}$. The amplitude of the  power spectrum decreases quickly as $\Delta\nu$ is increased from zero because the Bessel functions oscillate out of phase. For $l=500$, the amplitude falls to 1\% of the peak value when $\Delta\nu \approx 1.3$ MHz. For multipoles $l=1000$ and $l=5000$, the amplitude falls to 1\% of the peak value for $\Delta\nu \approx  0.8$ and $0.2$ MHz respectively. For comparison, the spectral resolution of the LOFAR experiment $\sim$ 10 kHz or better \cite{lofar_freq}.  The strong frequency dependence of the multi-frequency angular power spectrum may be used to separate the signal from the much larger foregrounds which are expected to be highly correlated over such small frequency ranges \cite{fore}.

\subsection{Variation with redshift.}

Let us now consider the variation of the amplitude of the power spectrum with redshift or equivalently, with frequency $\nu = \nu_{\rm 0} / (1+z)$. We consider here,  redshifts that are well separated from each other. We  assume that the signal has been separated from the background and plot the peak value $C_l(\nu, \Delta\nu=0)$ for each redshift $z$.

\begin{figure}[!ht]
\begin{center}
\scalebox{0.8}{\includegraphics{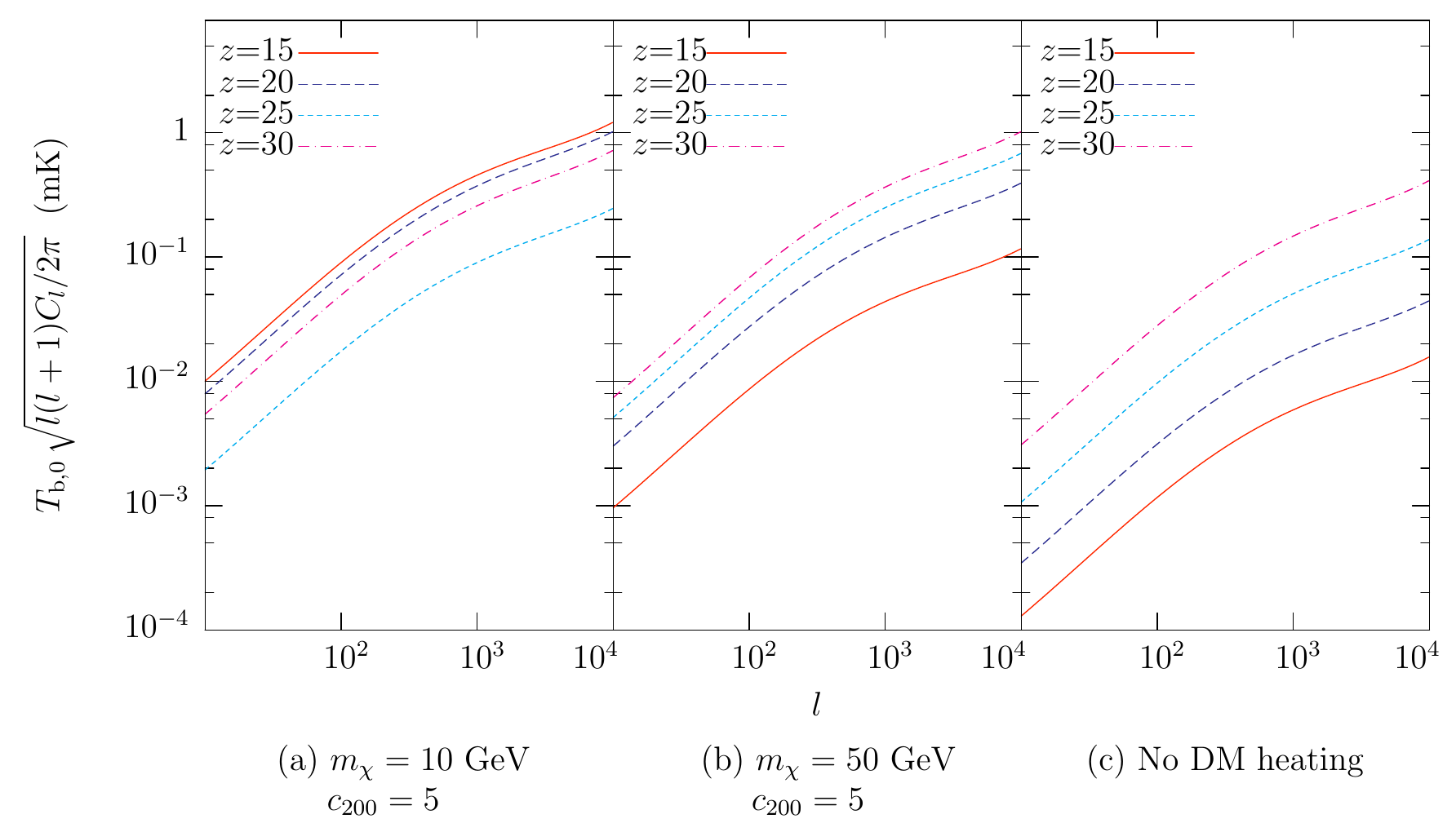}}
\end{center}
\caption{  Power spectrum $C_l(\nu, \Delta\nu=0)$ for different $z$. (a) and (b) show plots for $m_\chi = 10$ GeV and $50$ GeV respectively, with $c_{\rm 200} = 5$. (c) is plotted for the case of no heating by dark matter. In (a), the power spectrum decreases, and then increases, as we move from redshift $z=15$ to $30$, in contrast to (b) and (c). \label{fig11} }
\end{figure}

Fig. \ref{fig11} shows the H21 cm power spectrum $C_l(\nu, \Delta\nu=0)$ for redshifts $z=15,20,25$, and $30$. Plots (a) and (b) are shown for $m_\chi = 10$ and $50$ GeV respectively, for an assumed concentration parameter $c_{\rm 200} = 5$. (c) shows the case of no dark matter heating. In (a) the amplitude of the power spectrum of fluctuations \emph{decreases} as $z$ is increased from 15 to 25, and then increases from 25 to 30. The variation of $C_l$ with $z$ is due to the heating of the gas by dark matter annihilation. As the gas temperature increases, $T$ approaches $T_\gamma$, and hence the differential brightness temperature decreases, resulting in a loss of power. The power spectrum reaches a minimum when $T = T_\gamma$ and then increases again as $T > T_\gamma$ (the 21 cm background now appears in emission). This behavior is in contrast with (c) which shows a steady increase in power as the redshift is varied from 15 to 30. Thus observations of 21 cm fluctuations at different redshifts may be used to identify light dark matter models. The curves in (b) do not show a minimum with change in redshift. This is because for $c_{\rm 200} = 5$ and $m_\chi = 50$ GeV, the gas temperature $T$ never rises above the CMB temperature (see Fig. 4). Hence this effect is only visible with light dark matter models, or with large concentration parameters.

\begin{figure}[!ht]
\begin{center}
\scalebox{0.8}{\includegraphics{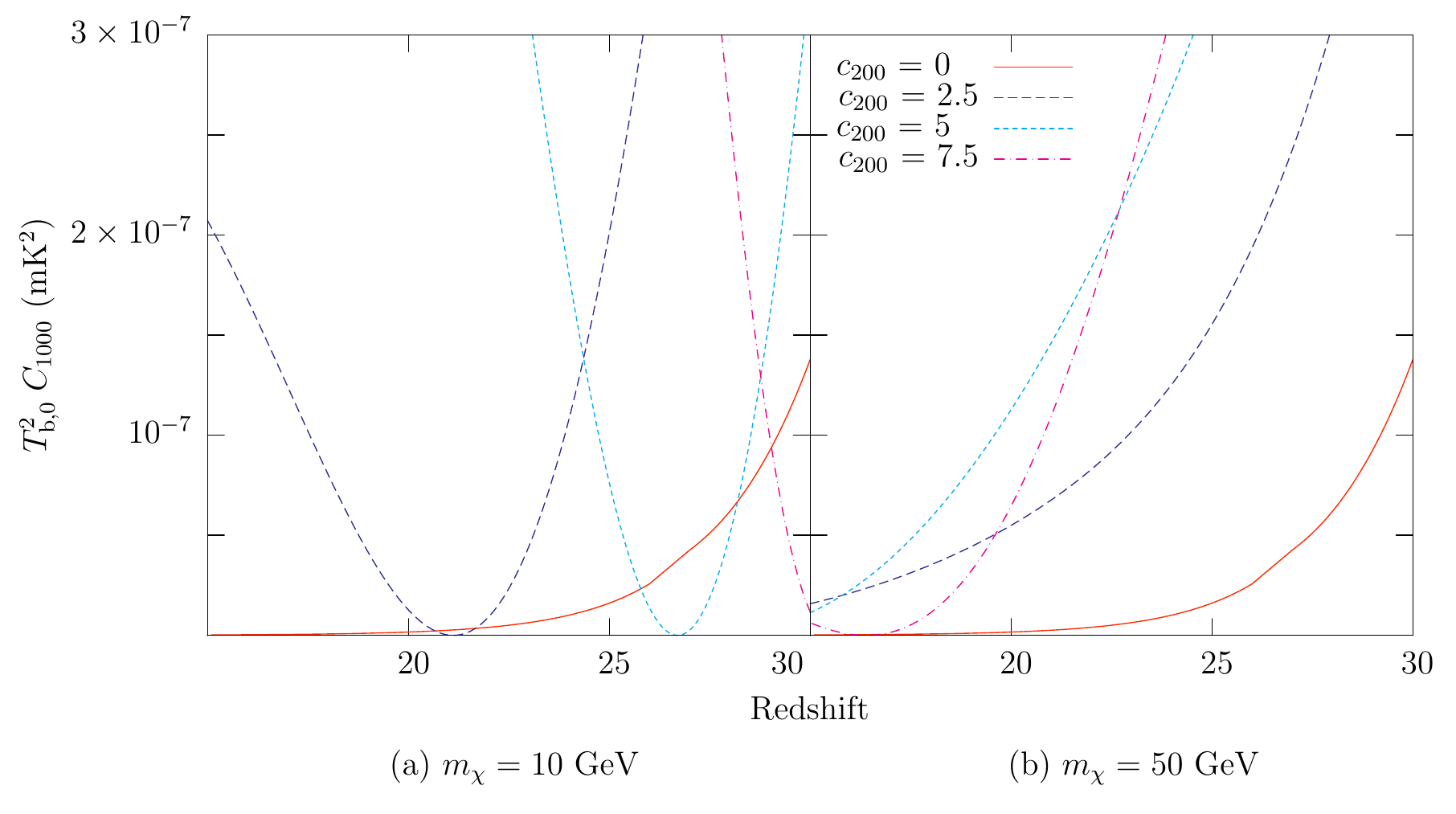}}
\end{center}
\caption{  The power spectrum $C_l(\nu, \Delta\nu=0)$ for $l=1000$ as a function of $z$, for $c_{\rm 200} = 2.5,5$, and 7.5. $\nu(z) = \nu_{\rm 0}/(1+z)$. Also shown is the case $c_{\rm 200} = 0$ which corresponds to no dark matter heating. (a) and (b) are plotted for $m_\chi = 10$ GeV and $50$ GeV respectively. \label{fig12} }
\end{figure}

Fig. \ref{fig12} shows the effect of variation with redshift, for $l=1000$, and for $c_{\rm 200} = 2.5,5$, and $7.5$. Also shown is the case $c_{\rm 200} = 0$ which does not take dark matter heating into account. (a) is plotted for $m_\chi=10$ GeV, while (b) shows $m_\chi = 50$ GeV. In (a), the power spectrum shows minima for $c_{\rm 200} = 2.5,5$, and 7.5. In contrast, in plot (b), we see a minimum (at $z = 16.4$) only with $c_{\rm 200} = 7.5$. Thus, for model (b) we would expect the power spectrum to show a minimum at this redshift when $m_\chi = 50$ GeV, $c_{\rm 200} = 7.5$. No minimum exists for $c_{\rm 200} =$ 2.5 or 5.

\section{Influence of astrophysical sources.}

So far, we have ignored the existence of stars and other luminous astrophysical sources, and considered heating solely due to dark matter annihilation. For redshifts $z<15$, this assumption is bound to break down as luminous astrophysical objects heat up and ionize the Universe. However, we expect that dark matter annihilation remains the dominant heating source at redshifts $z>20$. Let us now consider the effect of astrophysical sources on the brightness temperature.

\subsection{Region around early stars}

Let us first consider the formation of the earliest stars. These stars are massive, hot, metal-free, and form in dark matter halos with masses $M_{\rm halo} > 7 \times 10^5 M_\odot$ \cite{yoshi1}.  Each such star is surrounded by a hot, ionized region (radius $R_{\rm i}$) which in turn is surrounded by a warm neutral region (radius $R_{\rm n}$).  After the star has exhausted its fuel in a short time $\Delta z \approx 0.1 \; (\Delta t/$Myr) at $z=25$, the gas surrounding the star cools adiabatically, while the ionized region becomes partially neutral. In this simple estimate, we ignore the formation of supernovae, black holes, etc.

Let us consider stars with mass $M_{\rm star} \sim 100 \, M_\odot$ and lifetime $\sim$ few Myr.  The size and temperature of the ionized and neutral regions surrounding Pop.III stars have been estimated in \cite{chen}. Following \cite{chen}, we set a threshold temperature of $10^4$ K for the central ionized region up to $R_{\rm i}$ = 3 kpc, at an assumed formation redshift $z_{\rm s} = 25$. The gas temperature drops quickly to the cosmological average of $\approx 10$ K at $R_{\rm n}$ = 8 kpc. For simplicity, we approximate the temperature dependence on radius of \cite{chen} by a power law $T(r,z=25) \approx 10^4 \, \textrm{K} \; (R_{\rm i} / r)^{7.13}$. The differential brightness temperature averaged over the sphere surrounding a single star is
 \beq
 \langle T_{\rm b} \rangle = {\mathcal N} \int d\Omega \, \int ds \, T_{\rm b}(s,\theta),
 \eeq
where the integrals are over the solid angle subtended by a single sphere when observed by an instrument, and the line of sight passing through the sphere. ${\mathcal N}$ is a suitable normalization constant.  We have assumed here that single spheres are too small to be individually resolved. $\xi_{\rm c}$ and $\xi_\alpha$ are computed from Eq. (\ref{f}) and Eq. (\ref{xi_alpha_1}), with $E$ being the thermal energy. We find $\langle T_{\rm b} \rangle$ = +1.22 mK at $z=25$, increasing to a peak value of +1.41 mK at $z=22.7$, and decreasing thereafter. The low value of $\langle T_{\rm b} \rangle$ is due to the high degree of ionization of the central core and the rapid decrease in temperature away from the core. The temperature observed by an instrument = $f \, \langle T_{\rm b} \rangle + (1-f) \, T_{\rm other}$. $f$ is a ``filling fraction'' which takes into account that the individual spheres are in general, non-overlapping. $T_{\rm other}$ represents the contribution due to other sources. Using the star formation rate of \cite{yoshi2}, we estimate $f \sim \pi R^3_{\rm n} \, n_\ast(z=25) \approx 0.3$ where $n_\ast(z)$ is the number density of stars at $z$. Thus, if the dark matter is light ($m_\chi \sim 10$ GeV), we expect gas heating from early star formation to be only a small correction to the heating caused by particle annihilation, at high redshifts. Heavier dark matter particles ($m_\chi \sim 50$ GeV) with low concentration parameters $c_{\rm 200} \sim 5$ may be harder to distinguish from early stars. 

\subsection{Global evolution of the brightness temperature.}

The hot bubble framework discussed in the previous subsection is not applicable at late times, when the bubbles coalesce. Let us briefly discuss the astrophysical models considered by Pritchard and Loeb \cite{pl} who compute the global evolution of the 21 cm brightness temperature in the presence of astrophysical sources. Three models (named A,B, and C) are considered in \cite{pl} depending on the star formation efficiency, and the number of ionizing, Ly-$\alpha$, and X-ray photons produced by the sources. At a redshift $z=25$, the spin temperature is nearly equal to the CMB temperature and the brightness temperature relative to the CMB $\sim$ -2 mK for all models (see Fig. 1 of \cite{pl}). For smaller redshifts, the spin temperature starts to follow the gas temperature again due to increased Ly-$\alpha$ coupling, resulting in a minimum at $z\approx 17$ with a H21 cm differential brightness temperature $\approx -100$ mK for models A and B and $\approx$ -75 mK for model C. The ionized fraction is still very small. The differential brightness temperature passes through zero, becoming positive at $z \approx 14 - 15$, as the gas is heated above the CMB temperature. A local maximum is reached at $z \approx 12-14$, quickly falling off to zero at lower redshifts as the Universe reionizes.

    These models based on gas heating by luminous sources also exhibit a minimum in the amplitude of fluctuations at redshifts $z \approx 14-15$, when the gas temperature exceeds the CMB temperature (as long as $x_{\rm ion}$ is still small), as described in the previous section. Thus detecting a minimum in the power spectrum at redshifts $z<15$ cannot be interpreted as due to the presence of WIMP dark matter. For higher redshifts $z>20$, astrophysical sources have very little influence upon the global gas temperature and the average 21 cm differential brightness temperature. We therefore expect our results based on dark matter heating to remain valid at these redshifts. The existence of a minimum in the amplitude of fluctuations at redshifts $z>20$ would be a possible indication of heating by dark matter annihilation.

\section{Conclusions.}

We studied the effect of dark matter annihilation on the CMB polarization and the H21 cm power spectra, for different values of the particle mass, and assumed halo concentration parameter. We considered dark matter halos fitted with an NFW profile, and computed their luminosity. We used the Press-Schechter theory to determine the number of dark matter halos, and derived an expression for the energy absorbed by gas at a given redshift $z$. We compared this value to the energy generated at $z$ and found that only a small fraction of the generated energy is absorbed (Fig. 1, peak absorption $\sim$ 20\% for 10 GeV WIMPs, and $\sim$ 6\% for 100 GeV WIMPs). About $30\%$ of the absorbed energy goes into ionization and an equal amount into heating.  We also considered the effect of varying the minimum halo mass which is determined by the particle physics model (Fig. 2). We did not  include secondary ionizations, bremsstrahlung processes, etc, nor did we account for the increase in halo density with time. Thus, more accurate treatments could result in a larger absorption of energy. We computed the evolution of the ionized fraction (Fig. 3) and the gas temperature (Fig. 4) with redshift, for different values of the dark matter particle mass $m_\chi$ and assumed concentration parameter $c_{\rm 200}$. 

We then investigated the observable effects of ionization. We assumed a reionization model characterized by a redshift $z_\ast$ such that $x_{\rm ion} = 1$ for $z < z_\ast$. In models with dark matter annihilation, $x_{\rm ion} \ne 0$ for $z > z_\ast$. We compared these models with the standard reionization model and calculated the value of $z_\ast$ for different particle masses and concentration parameters, for an optical depth $\tau = 0.087$ (Table I). Figs. 5 and 6 show the reionization history and polarization power spectra for the model with $m_\chi = 1$ GeV, $c_{\rm 200} = 10$. The change in the polarization power spectrum is small, but observable for very light ($\sim 1$ GeV) dark matter particles, and for favorable halo parameters ($c_{\rm 200} \sim 10$). Future CMB experiments such as Planck and CMBPol may be able to distinguish the two scenarios, and hence distinguish light dark matter models from others. It will however be difficult to distinguish the dark matter reionization scenario from another model with gradual reionization.

We also studied the effect of dark matter annihilation on the Hydrogen 21 cm power spectrum. In Fig. 7, we compared the collisional and Ly-$\alpha$ coupling terms. Fig. 8 shows the differential brightness temperature of H21 cm radiation $T_{\rm b}$, for $m_\chi$ = 10 and 50 GeV and for concentration parameters $c_{\rm 200}$ = 5, 10, and 20. Also shown is the case of no dark matter heating ($c_{\rm 200} = 0$). We then considered perturbations to the baryon density ($\delta_{\rm n}$), ionized fraction ($\delta_{\rm x}$), and gas temperature ($\delta_{\rm T}$), and computed the multifrequency angular power spectrum of fluctuations $C_l(\nu,\Delta\nu)$ in $T_{\rm b}$.  Fig. 9 shows the time evolution and the relative amplitude of ionization and temperature fluctuations, assuming $\delta_{\rm x} = \delta_{\rm T} = \delta_{\rm n}$ initially. It was shown that ionization and temperature fluctuations are not very significant at low redshifts ($z < 30$). We studied the variation of the power spectrum $C_l(\nu,\Delta\nu)$ with frequency $\Delta\nu$. For $l=500$, $C_l \approx 0$ for $\Delta\nu \gtrsim$ 1 MHz, while for $l=5000$, $C_l$ falls off to $\approx 0$ for $\Delta\nu \gtrsim$ 0.1 MHz. The frequency dependence of $C_l(\nu,\Delta\nu)$ may be used to separate the Hydrogen 21 cm signal from the foreground contaminants. Fig. 11 shows the H21 power spectrum for different redshifts. Fig. 11(a) shows the case $m_\chi = 10$ GeV, $c_{\rm 200} = 5$. In this model, the amplitude of fluctuations decreases as $z$ is increased from 15 to 25, and then increases from 25 to 30. This behavior is due to the gas temperature $T$ approaching the CMB temperature $T_\gamma$ and then increasing beyond $T_\gamma$. Fig. 11(b) ($m_\chi = 50$ GeV, $c_{\rm 200} = 5$) does not show a minimum at any redshift, and is a monotonous function of $z$. This is because for this model, the gas temperature $T$ never exceeds the CMB temperature at any redshift in the considered range (compare Fig. 4(a) with Fig. 4(b)). Fig. 11(c) shows the power spectrum for the standard scenario with no dark matter heating.  Fig. 12 shows the variation of the power spectrum with redshift in the range 15 to 30, for $l=1000$, for different values of $c_{\rm 200}$. In (a), the power spectrum shows a minimum with redshift for all values of $c_{\rm 200} \ne 0$ (compare with Fig. 11(a)). In (b), the models with $c_{\rm 200} = 2.5$ and $5$ show no minimum (compare with Fig. 11(b)), while the model with $c_{\rm 200} = 7.5$ shows a minimum at $z=16.4$. Future Hydrogen 21 cm experiments such as LOFAR and SKA may be able to distinguish light dark matter models ($m_\chi \sim 10$ GeV) from the standard scenario by measuring the power spectrum at different redshifts. However, it would be difficult to distinguish models with larger dark matter masses ($m_\chi \sim 50$ GeV) from the standard scenario, except for favorable halo parameters ($c_{\rm 200} > 10$). In certain dark matter models where the dark matter particle is produced by a non-thermal mechanism, the annihilation rate can be much larger than in the standard scenario considered here. In such cases, the ionization and heating rates are substantially higher \cite{hooper} which may make it easier to distinguish the effects of dark matter annihilation from other sources.

Finally, we discussed the effect of astrophysical sources on our calculations. We found that at high redshifts, heating by early stars is not a very large effect, especially for light dark matter particles. Later star formation at lower redshifts, and the formation of the first galaxies will likely be larger contaminants. We discussed the evolution of the brightness temperature at lower redshifts in the models of \cite{pl}. Heating by astrophysical sources also results in a minimum in the amplitude of fluctuations at redshifts $z < 15$.  However for redshifts $z > 20$, we expect dark matter annihilation to be the relevant effect. Thus detecting a minimum in the amplitude of Hydrogen 21 cm fluctuations at $z>20$ would be a possible indication of heating by dark matter annihilation.

\acknowledgments{The authors would like to thank the anonymous referee for many comments which helped improve the manuscript. We thank Zolt\'{a}n Haiman, Glenn Starkman, and Rennan Barkana, and Marco Cirelli for helpful discussions. This work was supported by the Deutsche Forchungsgemeinschaft under grant GRK 881. }

\end{document}